\documentclass[10pt,aip,sdy,superscriptaddress,floatfix,longbibliography,reprint]{revtex4-2}
\usepackage{graphicx}
\usepackage{dcolumn}
\usepackage{bm}
\usepackage{amsmath}
\usepackage{subfigure}
\usepackage{url} 
\usepackage{hyperref} 
\usepackage{xcolor}

\begin{document}

\title{Transient grating spectroscopy on a DyCo$_5$ thin film with femtosecond extreme ultraviolet pulses}

\author{Victor Ukleev}
\email{victor.ukleev@helmholtz-berlin.de}
\affiliation{Helmholtz-Zentrum Berlin f\"ur Materialien und Energie, D-12489 Berlin, Germany}
\author{Ludmila Leroy}
\affiliation{Swiss Light Source, Paul Scherrer Institute, 5232 Villigen PSI, Switzerland}
\author{Riccardo Mincigrucci}
\affiliation{Elettra - Sincrotrone Trieste S.C.p.A., 34149 Basovizza, Trieste, Italy}
\author{Dario Deangelis}
\affiliation{Elettra - Sincrotrone Trieste S.C.p.A., 34149 Basovizza, Trieste, Italy}
\author{Danny Fainozzi}
\affiliation{Elettra - Sincrotrone Trieste S.C.p.A., 34149 Basovizza, Trieste, Italy}
\author{Nupur Ninad Khatu}
\affiliation{Elettra - Sincrotrone Trieste S.C.p.A., 34149 Basovizza, Trieste, Italy}
\affiliation{Department of Molecular Sciences and Nanosystems, Ca' Foscari University of Venice, Scientific Campus, Via Torino 155, 30172 Mestre (Venice), Italy}
\affiliation{European X-ray Free Electron Laser, Holzkoppel 4, 22869 Schenefeld, Germany}
\author{Ettore Paltanin}
\affiliation{Elettra - Sincrotrone Trieste S.C.p.A., 34149 Basovizza, Trieste, Italy}
\author{Laura Foglia}
\affiliation{Elettra - Sincrotrone Trieste S.C.p.A., 34149 Basovizza, Trieste, Italy}
\author{Filippo Bencivenga}
\affiliation{Elettra - Sincrotrone Trieste S.C.p.A., 34149 Basovizza, Trieste, Italy}
\author{Chen Luo}
\affiliation{Helmholtz-Zentrum Berlin f\"ur Materialien und Energie, D-12489 Berlin, Germany}
\author{Florian Ruske}
\affiliation{Helmholtz-Zentrum Berlin f\"ur Materialien und Energie, D-12489 Berlin, Germany}
\author{Florin Radu}
\affiliation{Helmholtz-Zentrum Berlin f\"ur Materialien und Energie, D-12489 Berlin, Germany}
\author{Cristian Svetina}
\affiliation{SwissFEL, Paul Scherrer Institute, 5232 Villigen PSI, Switzerland}
\affiliation{Madrid Institute for Advanced Studies, IMDEA Nanociencia, Calle Faraday 9, Ciudad Universitaria de Cantoblanco, Madrid, 28049, Spain}
\author{Urs Staub}
\email{urs.staub@psi.ch}
\affiliation{Swiss Light Source, Paul Scherrer Institute, 5232 Villigen PSI, Switzerland}

\date{\today}

\begin{abstract}
Surface acoustic waves (SAWs) are excited by femtosecond extreme ultraviolet (EUV) transient gratings (TGs) in a room-temperature ferrimagnetic DyCo$_5$ alloy. TGs are generated by crossing a pair of EUV pulses from a free electron laser (FEL) with the wavelength of 20.8\,nm matching the Co $M$-edge, resulting in a SAW wavelength of $\Lambda=44$\,nm. Using the pump-probe transient grating scheme in a reflection geometry the excited SAWs could be followed in the time range of -10 to 100\,ps in the thin film. Coherent generation of TGs by ultrafast EUV pulses allows to excite SAW in any material and to investigate their couplings to other dynamics such as spin waves and orbital dynamics. In contrast, we encountered challenges in detecting electronic and magnetic signals, potentially due to the dominance of the larger SAW signal and the weakened reflection signal from underlying layers. A potential solution for the latter challenge involves employing soft X-ray probes, albeit introducing additional complexities associated with the required grazing incidence geometry.
\end{abstract}

\maketitle

\section{Introduction}

In recent years, significant advancements have been made in transient grating (TG) spectroscopy techniques, leveraging the capabilities of extreme ultraviolet (EUV) and x-rays delivered by free-electron lasers (FELs) \cite{bencivenga2015four,bencivenga2019nanoscale,rouxel2021hard}. Recent developments in FEL photon sources have enabled new approaches for probing ultrafast dynamics at the nanoscale, such as TGs, which was recently used to study thermoelastic and magnetic dynamics in condensed matter systems \cite{maznev2018generation,maznev2021generation,rouxel2021hard,milloch2021nanoscale,bencivenga2019nanoscale,svetina2019towards,weder2020transient,rouxel2021hard,ksenzov2021nanoscale,yao2022all}.

In EUV TG, coherent FEL beams intersecting at the sample surface give rise to spatially periodic excitation patterns with periods in the 10s of nm range and time duration on the order of tens of femtoseconds \cite{eichler2013laser}. This allows for exploring dynamics at both nanoscale dimensions and ultrafast timescales.

EUV TGs can be used for studying nanoscale magnetic dynamics, provided that the probe is tuned to a magnetic edge, as recently demonstrated in a Gd-Co alloy at the Co $M$-edge \cite{ksenzov2021nanoscale}. This element-specific tool can help in shedding light on the coupling of magnetic and structural dynamics, inherently intertwined through magnetoelastic interactions \cite{januvsonis2016transient}.

\begin{figure*}
\begin{center}
\includegraphics[width=1\linewidth]{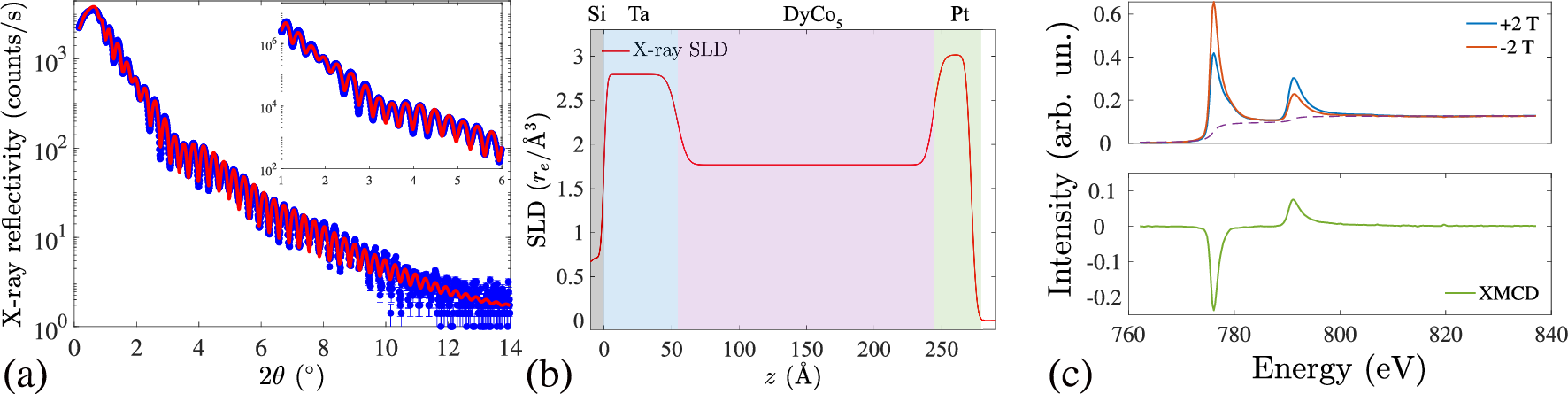}
        \caption{(a) Measured (symbols) and fitted (solid line) x-ray reflectometry (XRR) curves from [Ta (5 nm)/DyCo$_5$ (20 nm)/Pt (2 nm)] sample on a Si substrate. The inset is the zoom-in part of the XRR curve, showing the excellent agreement between the measured data and the fitted model. (b) Reconstructed scattering length density (SLD) depth profile obtained from the XRR model in the units of the classical electron radius $r_e=2.81794\cdot10^{-5}$\,\AA. The coordinate $z=0$\,\AA~corresponds to the Si / Ta interface. (c) X-ray absorption spectroscopy (XAS, top panel) and x-ray magnetic circular dichroism (XMCD, bottom panel) spectra measured from the same sample at room temperature and applied magnetic field of $\pm2$\,T by means of the total electron yield. The dashed line in the XAS plot is the step function used in order to correct the spectra for the sum rule analysis.}
        \label{fig1}
\end{center}
\end{figure*}

The binary intermetallic DyCo$_5$ alloy belongs to the family of binary rare-earth (RE) - transition metal (TM) compounds, renowned for their magnetic properties that find applications in permanent magnet technologies \cite{coey2020perspective}, spintronics \cite{gonzalez2021applied}, and ultrafast optical switching \cite{hansen1991magnetic}. Particularly, DyCo$_5$ exhibits remarkable magnetic properties including a high Curie temperature of approximately 970\,K, attributable to the large exchange coupling of Co spins, as well as a high coercive field stemming from a considerably high magnetic anisotropy \cite{andreev1991thermal,radwanski1987origin}. Such properties arise from the antiferromagnetic coupling between the Dy and Co sub-lattices, resulting in a nontrivial magnetic phase diagram which includes a compensation point at 174\,K and a spin-reorientation transition at approximately $T \sim 360$\,K \cite{tie1991magnetic,donges2017magnetization,grechishkin2017domain}. The presence of a compensation point, a key factor for thermally-assisted magnetization switching \cite{kirilyuk2010ultrafast,radu2011transient} making DyCo$_5$ a promising candidate for ultrafast optical manipulation. Moreover, thin films and nanostructures of ferrimagnetic Dy-Co alloys combine magnetic properties, such as perpendicular magnetic anisotropy and room-temperature magnetization, which are relevant for applications in magnetic memory storage \cite{radu2012perpendicular,unal2016ferrimagnetic,radu2018ferrimagnetic} and spintronic devices \cite{chen2020observation,seifert2021frequency}.

RE-TM alloys are featured by strong magnetoelastic coupling \cite{del1987magnetoelastic,kamarad1995magnetic}, making them suited for exploring the interaction between the structural and magnetic dynamics induced by EUV TGs. In the present study we used EUV TG in reflection geometry with the probe tuned to the Co $M$-edge to detect nanoscale magnetic and thermoelastic dynamics on the surface of a DyCo$_5$ sample. The analysis revealed a predominance of thermoelastic dynamics driven by surface acoustic waves (SAW). While no discernible signal from magnetization dynamics was observed. This suggests that the signal associated with the magnetization grating either falls below the detection limit or is dominated by the robust signal generated by coherent surface displacements. Consequently, it is inferred that EUV TG in reflection geometry may not be the optimal choice for detecting magnetic dynamics. Nevertheless, the generation and detection of SAWs in thin DyCo$_5$ films hold intrinsic importance. SAWs at nanoscale wavelengths possess various applications in spintronics \cite{puebla2022perspectives}, and the application of EUV TG allows access to this wavelength range without the need for tailored nanostructures on the sample.

\section{Materials and Methods}

\subsection{Synthesis}

The 20\,nm-thick DyCo$_5$ film was grown onto a Si substrate with a 5\,nm-thick Ta buffer by means of magnetron sputtering at room temperature under Ar atmosphere of $1.5\cdot 10^{-3}$\,mbar (with a base pressure of $10^{-8}$\,mbar) using the MAGSSY chamber at the Helmholtz-Zentrum Berlin (HZB, Germany). The correct stoichiometry of the alloy was achieved by varying the deposition rate of Co and Dy targets in the co-evaporation scheme. The film was capped by 2\,nm of Pt to prevent surface oxidation. More details of the sample growth are given in Ref. \onlinecite{luo2019x}.

\begin{figure*}
\begin{center}
\includegraphics[width=1\linewidth]{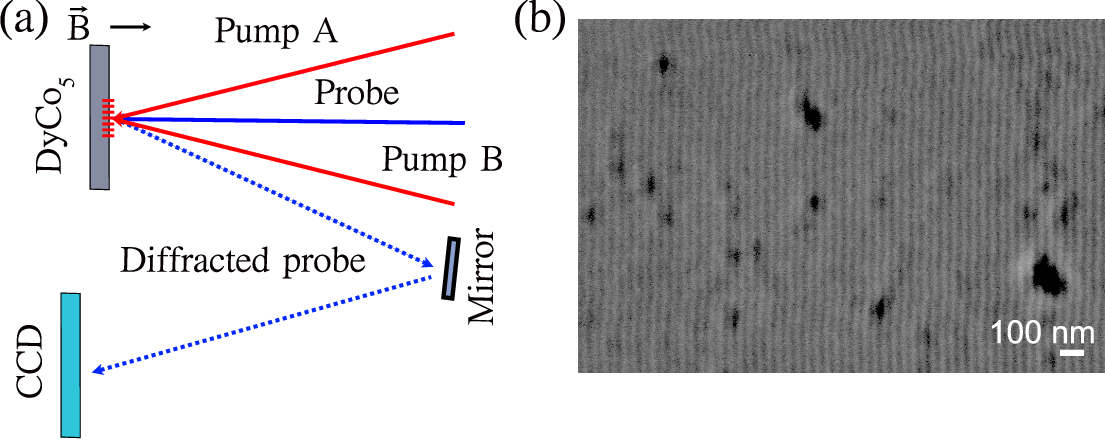}
        \caption{(a) Sketch of the EUV TG experimental geometry. Circularly polarized extreme ultraviolet (EUV) pump beams A and B (red arrows) are crossed at the sample generating the interference pattern. The probe beam (full blue arrow) transiently diffracted off this pattern in reflection geometry (dashed blue arrow) is the EUV TG signal, which is routed to the CCD detector by the multilayer mirrors. (b) Scanning electron microscopy image of the permanent grating with periodicity of 44\,nm permanently generated on the sample surface by the crossed pump beams with the total fluence of $40$\,mJ/cm$^2$.
        %(c) Spatio-temporal dynamics of the sample before and after the generation of the TG at the time-zero point. The response is visualized using the sinusoidal spacial periodicity of the TG shown in the panel (b), and the measured time traces of the diffracted signal.
        }
        \label{fig2}
\end{center}
\end{figure*}

\subsection{X-ray reflectometry}

Due to the limited penetration depth of EUV radiation in the material, experiments were performed in reflection geometry, which demands a high-quality (uniform and low roughness) surface of the sample. To assess the surface quality, a characterization of the DyCo$_5$ film was performed using x-ray reflectometry (XRR) (Figure \ref{fig1}). The XRR measurements were conducted at the X-ray CoreLab facility of the HZB using the PANalytical MPD instrument. The experimental data were obtained in the specular reflection geometry using x-rays with a wavelength $\lambda=1.54$\,\AA~and fitted employing the GenX 3 software \cite{glavic2022genx} for precise assessment of the surface and the interlayer roughness.

Oscillations in the XRR curve, known as Kiessig fringes, were observed up to a high $2\theta$ angle of 12$^\circ$, indicative of the high quality of the film \cite{daillant2008x} (Figure \ref{fig1}a). Fitting the XRR data yielded insights into the structural quality of individual layers (including their thickness and roughness) and enabled reconstruction of the in-depth electronic scattering length density (SLD) profile \cite{daillant2008x}. Figure \ref{fig1}a shows a good agreement between XRR data and the best fit curve, which is achieved by assuming well-defined layers of Ta, DyCo$_5$, and Pt on the Si substrate, each one with densities close to their nominal values. The corresponding in-depth ($z$) SLD distribution within the sample is displayed in Figure \ref{fig1}b. The model closely matches the anticipated layer thicknesses, while also suggesting sub-nanometer roughness at the sample surface. This finding attests a good sample quality, thus laying the ground for successful FEL experiments in the EUV regime.

\subsection{X-ray magnetic circular dichroism}

The magnetic moment of the sample was probed utilizing soft x-ray magnetic circular dichroism (XMCD) in the total electron yield (TEY) detection mode. The probing depth of the TEY method extends to a few nanometers from the surface \cite{stohr2006magnetism}, thereby providing valuable insights into the surface quality and potential oxidation state of the sample. X-ray absorption spectra (XAS) were acquired with a 77\% degree of right-circular polarization using the VEKMAG instrument at BESSY-II, Berlin, Germany \cite{noll2016mechanics}. Given that the TG experiment was conducted using EUV radiation at the Co $M$ edge, our focus is limited to the soft x-ray spectroscopy characterization of cobalt. To extract the magnetic contribution from the XAS data, spectra were measured under saturating magnetic fields of $\pm2$\,T at the Co $L_{2,3}$ edges. The XAS and XMCD spectra are depicted in Figure \ref{fig1}c.

The absence of multiplet features in the XAS spectra indicate the metallic nature of the DyCo$_5$ film and the absence of surface oxidation. The XMCD signal was observed at both $L_{2,3}$ edges (bottom panel in Figure \ref{fig1}), indicating the fully polarized ferromagnetic state of cobalt at room temperature. Employing sum rule analysis on the XMCD data, the extracted spin and orbital magnetic moments of cobalt are found to be $m_\textrm{s}=1.20(3)\mu_\textrm{B}$ and $m_\textrm{l}=0.16(1)\mu_\textrm{B}$, respectively. In agreement with the previous study \cite{luo2019x}, the net moment of Co at the DyCo$_5$ surface probed by TEY is considerably reduced compared to the bulk of the film ($m_\textrm{s}+m_\textrm{l} \approx 1.6\mu_\textrm{B}$) as seen in transmission and fluorescence XMCD measurements \cite{donges2017magnetization,luo2019x}. 

\subsection{X-ray transient grating spectroscopy}

\begin{figure*}
\begin{center}
\includegraphics[width=1\linewidth]{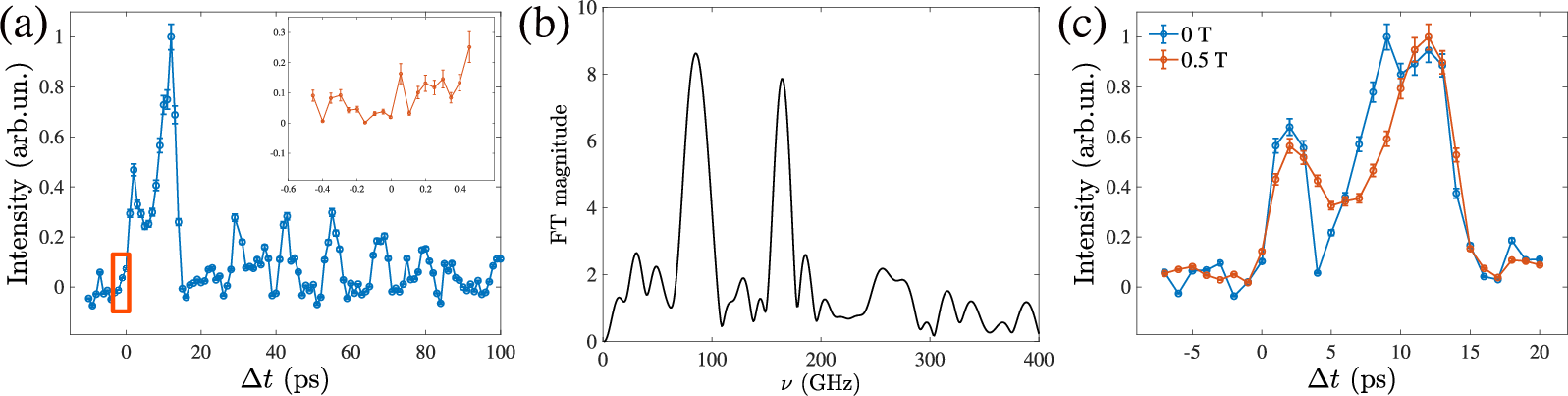}
        \caption{(a) Time trace of the EUV TG signal. The inset shows the high temporal resolution scan in the ultrafast region ($\pm 0.5$\,ps) where the electronic signal could be expected. Error bars have been calculated using the standard deviation of the signal at negative time delays.} (b) Zero-padded Fourier transform of the 20-100 ps part of the spectrum. (c) TG signals measured at zero magnetic field and at $B=0.5$\,T. The two waveforms were collected in different locations in the sample surface. While the magnitude of the first two oscillations and the signal-to-noise ratio differ from spot to spot, the time dependence of the oscillations at $\Delta t > 15$\,ps is independent on the sample position.
        \label{fig3}
\end{center}
\end{figure*}

The TG experiments were conducted at the EIS-TIMER beamline at the FERMI FEL, Trieste, Italy \cite{mincigrucci2018advances,bencivenga2019nanoscale}. The experimental configuration is described in Refs. \onlinecite{maznev2021generation,mincigrucci2018advances} and schematically illustrated in Figure \ref{fig2}a, including a variable magnetic field ($B$) applied to the sample, whose value, was varied in the 0 to 0.5\,T range by adjusting the distance between the sample and a permanent magnet.

Two circularly polarized pump pulses, designated as Pump A and Pump B in Figure \ref{fig2}a, intersected at the sample surface under an angle of $\theta=27.6^\circ$ to generate an interference pattern with a period of $\Lambda=\lambda / 2  \sin (\theta)=44$\,nm, where $\lambda=20.8$\,nm is the excitation wavelength. The time-delayed vertically polarized probe pulse at the same wavelength ($\lambda_\textrm{pr}=20.8$\,nm; matching the Co $M$-edge) was directed onto the sample at an angle of 13$^\circ$. In this configuration, the vertically polarized beam is mainly sensitive to the out-of-plane magnetization component \cite{paolasini2014resonant}, which, in DyCo$_5$ films at room temperature, corresponds to the magnetization direction \cite{radu2012perpendicular}.

The probe beam, back-diffracted from the TG, was then reflected by the multilayer mirror onto an in-vacuum two-dimensional charge-coupled device (CCD) detector, using the same setup described in Refs. \cite{maznev2021generation,mincigrucci2018advances}.

The total fluence of the two pump beams at the sample was 10\,mJ/cm$^2$. For each time delay, 1000 FEL shots were accumulated to measure the TG intensity. To mitigate the effects of radiation damage, each time trace was acquired at a different location on the sample.

By increasing the FEL fluence above 20\,mJ/cm$^2$, a permanent grating was imprinted onto the film, which involves structural and magnetic changes of the sample \cite{ukleev2023effect}. The characteristics of the permanent grating were assessed through scanning electron microscopy (SEM) utilizing a Zeiss Merlin instrument at the Correlative Microscopy and Spectroscopy CoreLab of the HZB.

The period of the grating observed in SEM, $\Lambda=44\pm1$\,nm, closely aligns with the value calculated from the crossing angle and wavelength of the two pump beams (Figure \ref{fig2}b) \cite{naumenko2019thermoelasticity}.

\section{Results and discussion}

Time-dependent EUV TG signals were acquired by scanning the delay between the EUV TG and probe beams $\Delta t$ in increments of 1\,ps. Moreover, to encompass the eventual ultrafast electronic and magnetic responses, a detailed investigation of the first picosecond time window (-0.5 to 0.5\,ps) was conducted, with a step size of 0.1\,ps.

%The spatio-temporal signal originating from the TG-induced excitation in the sample is presented in Figure \ref{fig2}c. This visualization merges the spatial grating excitation with period $\Lambda$ and the recorded time evolution of the TG signal. Further elaboration on the experimental time traces is provided in the subsequent sections.

Upon sample excitation, the application of EUV TG creates a spatially periodic temperature distribution across the surface of the DyCo$_5$ film, with a period $\Lambda$. This excitation triggers the rapid thermal expansion and consequent ultrafast generation of counter-propagating SAWs, as described in Ref. \onlinecite{rogers2000optical}. 

As depicted in Figure \ref{fig3}a, the normalized diffracted intensity, as a function of $\Delta t$, exhibits discernible periodic oscillations with the period of 12\,ps.

The zero-padded Fourier transform of the time trace in the range $\Delta t = 20-100$ ps, where the signal exhibits more regular oscillations, reveals the presence of SAWs. These waves are characterized by two principal phonon harmonics at frequencies $\nu=84$ GHz and 166 GHz, as depicted in Figure \ref{fig3}b.

The lower frequency corresponds to the SAW with a phase velocity of 3700 m/s, comparable to SAW velocities measured in TbCo multilayers \cite{zhou2014multilayer}. The double of the frequency can be attributed to the quadratic dependence of the diffraction intensity $I$ on the surface displacement \cite{maznev2021generation}:
\begin{eqnarray*}
I (t) &\sim& (S_\mathrm{th}(t) - A \exp{(-\frac{t}{\tau})} \cos{(\omega t}))^2=\\ &&
S_\mathrm{th}(t)^2 + 2 S_\mathrm{th}(t) A \exp{(-\frac{t}{\tau})} \cos{(\omega t}) + \\ &&
\frac{1}{2} A^2 \exp{(-\frac{2t}{\tau}) (\cos{(2\omega t})+1)}. 
\end{eqnarray*}
Here, $S_\mathrm{th}(t)$ describes the exponential decay of the thermal grating \cite{kading1995transient}, while $A$, $\tau$, and $\omega=2\pi \nu$ are, respectively, the amplitude, decay time, and angular frequency of the SAW. 
At $\Lambda$=44\,nm, the SAW oscillations most likely persist without decay within a 100\,ps timeframe, and the observed signal corresponds to the gradual intensity decline of the thermal grating, modulated by the SAW.

Surprisingly, within the initial 1 ps time window (as illustrated in the inset of Figure \ref{fig3}a), the ultrafast electronic response from cobalt remains conspicuously absent, as well as the rise of the magnetic response -- an observation that contrasts the Gd-Co study conducted in transmission geometry, where electronic and magnetic dynamics are observed within 500 fs after the FEL pulse. This initial response is subsequently followed by a decaying TG signal spanning tens of picoseconds. The magnetic TG signal in the Gd-Co study became evident when comparing data acquired in magnetically saturated and remanent states: the magnetic response was only discernible in the former scenario, attributed to the emergence of multiple magnetic domains in zero field \cite{ksenzov2021nanoscale}. However, in the present study, no significant response to external magnetic fields is observed. In DyCo$_5$, applying a saturation magnetic field of 0.5 T does not cause any noticeable change in dynamics, as shown in Figure \ref{fig3}c.

The absence of any magnetic appreciable response in the present EUV study is surprising, since a previous study has demonstrated ultrafast demagnetization via 40 fs infrared (IR) laser pulses with a photon energy of 1.5 eV. At incident laser fluences of a similar magnitude, i.e. 10-15 mJ/cm$^2$, IR photons induced 40\% and 80\% demagnetization of the Co sub-lattice within the first 2 ps \cite{abrudan2021element}. On the other hand, an ultrafast magnetic response to the both IR and EUV pumps was observed in Gd-Co, a system rather similar to DyCo$_5$. The absence of the magnetic TG signal in DyCo$_5$ thus deserves further discussion. Indeed, it's noteworthy that even a modest increase in the total pump fluence is sufficient to induce the permanent imprint of the grating via radiation damage after about 20 s exposure. Hence, the pump fluence should also be sufficient to induce the change in magnetization.

Assuming that the fluence level was adequate to drive a magnetization grating, the absence of a detectable signal could be due to the experimental geometry. Indeed, the observed backward-diffracted signal arises from both the spatially periodic perturbation of the material's refractive index, which modulates the EUV reflectivity, and from coherent surface displacements arising from the thermal expansion driven by the thermal grating \cite{bencivenga2023extreme}. The latter contribution is expected to be relatively weaker in experiments conducted in transmission geometry \cite{ksenzov2021nanoscale} on sample with optimal thickness, since the signal from refractive index modulations increases on crossing the excited thickness of the sample. When the probe resonates with a magnetic edge, refractive index alterations come from electronic population modulation affecting core-hole transitions (electronic grating signal), dichroic component modulation (magnetization grating signal), and density modulation (density grating signal). The latter, featured by longitudinal acoustic phonons but typically weaker due to its dependence on the square of thermal expansion, contrasts with the more pronounced magnetization grating signal, as seen in previous studies \cite{ksenzov2021nanoscale,yao2022all}. The transmission geometry is thus advantageous in removing the competing signal due to the SAWs, making it in capturing the transient magnetic signal compared to reflection geometry, despite the high structural and magnetic quality of our sample. In addition, by taking into account absorption of Pt, DyCo$_5$ and Ta at the probe wavelength \cite{thompson2001x}, the transmission of the corresponding layers are 70\%, 22\% and 64\%, respectively. The backscattered intensity from vacuum/Pt, Pt/DyCo$_5$, DyCo$_5$/Ta, and Ta/Si interfaces are estimated as 2\%, 0.98\%, 0.016\%, and 0.0023\% of the intensity upstream the sample, respectively, as we estimate by taking into account reflectivity of each interface, and absorption of the incoming and reflected beams. Hence, the observed scattering mainly originates from the Pt/vacuum interface, which is expected to be not sensitive to the magnetic state of DyCo$_5$ when the probe wavelength is away from Pt edges, whilst intensities from DyCo$_5$/Ta and Ta/Si interfaces are two order of magnitude weaker. Therefore, TG experiments in reflection geometry require not only sub-nm roughness of surface and interfaces for ensuring high reflectivity, but also careful selection of the adjacent layers of the material under study. An alternative approach for isolating in reflection geometry a weak magnetic signal out of a strong SAW signal is to use transient polarization gratings, which is possible by employing a special FEL configuration \cite{foglia2023nanoscale}. Using $L$-edge soft x-ray energies to detect the magnetic TG signal could mitigate the issue of absorption losses. However, this approach complicates the experimental geometry since substantial soft X-ray reflectivity is only possible at grazing incidence \cite{foglia2023extreme,bencivenga2023extreme}.

\section{Conclusion}

This paper presents a study of the coherent generation of SAWs within DyCo$_5$ thin films through the utilization of femtosecond EUV FEL pulses in a transient grating geometry. SAWs hold significant promise for applications in spintronics, offering potential avenues for, e.g., coherent domain wall nucleation, topological structure manipulation \cite{yokouchi2020creation,chen2021surface,puebla2022perspectives}, and spin current generation \cite{kobayashi2017spin}.

Despite the high quality of the sample used in our study, we did not observe any electronic or magnetic signals. This could potentially be attributed to the significantly larger signal originating from SAWs and the complexities associated with reduces backscattering signal from relevant interfaces within the material. Our findings highlight the challenges encountered when employing EUV TG in reflection geometry, particularly in the context of detecting magnetic dynamics. While this setup potentially expands the applicability of the technique to a wider range of magnetic materials that may not be transparent to EUV radiation, it is essential to carefully consider these factors to ensure the reliable detection of magnetic phenomena. Addressing these challenges will be crucial for advancing our understanding of magnetic dynamics using EUV TG in reflection geometry.

\section*{Acknowledgements}

Authors thank FERMI free-electron laser facility for provided beamtime according to the proposal 20214012. The x-ray spectroscopy experiment was carried out at the beamline PM-2 VEKMAG at BESSY II synchrotron as a part of the proposal 231-11957. We kindly acknowledge X-ray and Correlative Microscopy and Spectroscopy CoreLabs of HZB for provided instrumentation and Ren\'e Gunder for his assistance with the sample characterization. We thank C. David, I. Bykova, J. Raabe, R. Abrudan and N. Jaouen for fruitful discussions. This study was supported by the Swiss National Science Foundation (SNSF), grants no. 200021\_165550/1, no. 200021\_169017 and and No. 200021\_196964  as well as the SNSF National Centers of Competence in Research in Molecular Ultrafast Science and Technology (NCCR MUST-No. 51NF40-183615). The research leading to these results has received funding from LASERLAB-EUROPE (grant agreement no. 871124, European Union’s Horizon 2020 research and innovation programme). Authors acknowledge financial support of the VEKMAG end station by the German Federal Ministry for Education and Research (BMBF 05K10PC2, 05K10WR1, 05K10KE1) by HZB. This project has received funding from the European Union's Horizon 2020 research and innovation program under the Marie Skłodowska-Curie Grant Agreement No. 860553.

\section*{Author Declarations}
\subsection*{Conflict of Interest}
The authors have no conflicts to disclose.
\subsection*{Author Contributions}
V.U., L.L., R.M., D.D., D.F., N.N.K, E.P., L.F., F.B. performed FEL experiments, C.L. and F.Ra. synthesized the sample, V.U., C.L., F.Ru., F.Ra characterized the sample, V.U. and R.M. analyzed the data, V.U., R.M., F.B., U.S. wrote the paper, V.U., R.M., C.S., U.S. jointly conceived the project.
\section*{Data availability Statement}
Sample characterization and FEL data is available from V.U. upon request.  

%aipnum4-2.bst 2019-01-14 (MD) hand-edited version of apsrev4-1.bst
\providecommand{\noopsort}[1]{}\providecommand{\singleletter}[1]{#1}%


%aipnum4-2.bst 2019-01-14 (MD) hand-edited version of apsrev4-1.bst
\providecommand{\noopsort}[1]{}\providecommand{\singleletter}[1]{#1}%
\begin{thebibliography}{50}%
\makeatletter
\providecommand \@ifxundefined [1]{%
 \@ifx{#1\undefined}
}%
\providecommand \@ifnum [1]{%
 \ifnum #1\expandafter \@firstoftwo
 \else \expandafter \@secondoftwo
 \fi
}%
\providecommand \@ifx [1]{%
 \ifx #1\expandafter \@firstoftwo
 \else \expandafter \@secondoftwo
 \fi
}%
\providecommand \natexlab [1]{#1}%
\providecommand \enquote  [1]{``#1''}%
\providecommand \bibnamefont  [1]{#1}%
\providecommand \bibfnamefont [1]{#1}%
\providecommand \citenamefont [1]{#1}%
\providecommand \href@noop [0]{\@secondoftwo}%
\providecommand \href [0]{\begingroup \@sanitize@url \@href}%
\providecommand \@href[1]{\@@startlink{#1}\@@href}%
\providecommand \@@href[1]{\endgroup#1\@@endlink}%
\providecommand \@sanitize@url [0]{\catcode `\\12\catcode `\$12\catcode `\&12\catcode `\#12\catcode `\^12\catcode `\_12\catcode `\%12\relax}%
\providecommand \@@startlink[1]{}%
\providecommand \@@endlink[0]{}%
\providecommand \url  [0]{\begingroup\@sanitize@url \@url }%
\providecommand \@url [1]{\endgroup\@href {#1}{\urlprefix }}%
\providecommand \urlprefix  [0]{URL }%
\providecommand \Eprint [0]{\href }%
\providecommand \doibase [0]{https://doi.org/}%
\providecommand \selectlanguage [0]{\@gobble}%
\providecommand \bibinfo  [0]{\@secondoftwo}%
\providecommand \bibfield  [0]{\@secondoftwo}%
\providecommand \translation [1]{[#1]}%
\providecommand \BibitemOpen [0]{}%
\providecommand \bibitemStop [0]{}%
\providecommand \bibitemNoStop [0]{.\EOS\space}%
\providecommand \EOS [0]{\spacefactor3000\relax}%
\providecommand \BibitemShut  [1]{\csname bibitem#1\endcsname}%
\let\auto@bib@innerbib\@empty
%</preamble>
\bibitem [{\citenamefont {Bencivenga}\ \emph {et~al.}(2015)\citenamefont {Bencivenga}, \citenamefont {Cucini}, \citenamefont {Capotondi}, \citenamefont {Battistoni}, \citenamefont {Mincigrucci}, \citenamefont {Giangrisostomi}, \citenamefont {Gessini}, \citenamefont {Manfredda}, \citenamefont {Nikolov}, \citenamefont {Pedersoli} \emph {et~al.}}]{bencivenga2015four}%
  \BibitemOpen
  \bibfield  {author} {\bibinfo {author} {\bibfnamefont {F.}~\bibnamefont {Bencivenga}}, \bibinfo {author} {\bibfnamefont {R.}~\bibnamefont {Cucini}}, \bibinfo {author} {\bibfnamefont {F.}~\bibnamefont {Capotondi}}, \bibinfo {author} {\bibfnamefont {A.}~\bibnamefont {Battistoni}}, \bibinfo {author} {\bibfnamefont {R.}~\bibnamefont {Mincigrucci}}, \bibinfo {author} {\bibfnamefont {E.}~\bibnamefont {Giangrisostomi}}, \bibinfo {author} {\bibfnamefont {A.}~\bibnamefont {Gessini}}, \bibinfo {author} {\bibfnamefont {M.}~\bibnamefont {Manfredda}}, \bibinfo {author} {\bibfnamefont {I.}~\bibnamefont {Nikolov}}, \bibinfo {author} {\bibfnamefont {E.}~\bibnamefont {Pedersoli}}, \emph {et~al.},\ }\bibfield  {title} {\enquote {\bibinfo {title} {Four-wave mixing experiments with extreme ultraviolet transient gratings},}\ }\href@noop {} {\bibfield  {journal} {\bibinfo  {journal} {Nature}\ }\textbf {\bibinfo {volume} {520}},\ \bibinfo {pages} {205--208} (\bibinfo {year} {2015})}\BibitemShut {NoStop}%
\bibitem [{\citenamefont {Bencivenga}\ \emph {et~al.}(2019)\citenamefont {Bencivenga}, \citenamefont {Mincigrucci}, \citenamefont {Capotondi}, \citenamefont {Foglia}, \citenamefont {Naumenko}, \citenamefont {Maznev}, \citenamefont {Pedersoli}, \citenamefont {Simoncig}, \citenamefont {Caporaletti}, \citenamefont {Chiloyan} \emph {et~al.}}]{bencivenga2019nanoscale}%
  \BibitemOpen
  \bibfield  {author} {\bibinfo {author} {\bibfnamefont {F.}~\bibnamefont {Bencivenga}}, \bibinfo {author} {\bibfnamefont {R.}~\bibnamefont {Mincigrucci}}, \bibinfo {author} {\bibfnamefont {F.}~\bibnamefont {Capotondi}}, \bibinfo {author} {\bibfnamefont {L.}~\bibnamefont {Foglia}}, \bibinfo {author} {\bibfnamefont {D.}~\bibnamefont {Naumenko}}, \bibinfo {author} {\bibfnamefont {A.}~\bibnamefont {Maznev}}, \bibinfo {author} {\bibfnamefont {E.}~\bibnamefont {Pedersoli}}, \bibinfo {author} {\bibfnamefont {A.}~\bibnamefont {Simoncig}}, \bibinfo {author} {\bibfnamefont {F.}~\bibnamefont {Caporaletti}}, \bibinfo {author} {\bibfnamefont {V.}~\bibnamefont {Chiloyan}}, \emph {et~al.},\ }\bibfield  {title} {\enquote {\bibinfo {title} {Nanoscale transient gratings excited and probed by extreme ultraviolet femtosecond pulses},}\ }\href@noop {} {\bibfield  {journal} {\bibinfo  {journal} {Science Advances}\ }\textbf {\bibinfo {volume} {5}},\ \bibinfo {pages} {eaaw5805} (\bibinfo {year} {2019})}\BibitemShut {NoStop}%
\bibitem [{\citenamefont {Rouxel}\ \emph {et~al.}(2021)\citenamefont {Rouxel}, \citenamefont {Fainozzi}, \citenamefont {Mankowsky}, \citenamefont {R{\"o}sner}, \citenamefont {Seniutinas}, \citenamefont {Mincigrucci}, \citenamefont {Catalini}, \citenamefont {Foglia}, \citenamefont {Cucini}, \citenamefont {D{\"o}ring} \emph {et~al.}}]{rouxel2021hard}%
  \BibitemOpen
  \bibfield  {author} {\bibinfo {author} {\bibfnamefont {J.~R.}\ \bibnamefont {Rouxel}}, \bibinfo {author} {\bibfnamefont {D.}~\bibnamefont {Fainozzi}}, \bibinfo {author} {\bibfnamefont {R.}~\bibnamefont {Mankowsky}}, \bibinfo {author} {\bibfnamefont {B.}~\bibnamefont {R{\"o}sner}}, \bibinfo {author} {\bibfnamefont {G.}~\bibnamefont {Seniutinas}}, \bibinfo {author} {\bibfnamefont {R.}~\bibnamefont {Mincigrucci}}, \bibinfo {author} {\bibfnamefont {S.}~\bibnamefont {Catalini}}, \bibinfo {author} {\bibfnamefont {L.}~\bibnamefont {Foglia}}, \bibinfo {author} {\bibfnamefont {R.}~\bibnamefont {Cucini}}, \bibinfo {author} {\bibfnamefont {F.}~\bibnamefont {D{\"o}ring}}, \emph {et~al.},\ }\bibfield  {title} {\enquote {\bibinfo {title} {Hard x-ray transient grating spectroscopy on bismuth germanate},}\ }\href@noop {} {\bibfield  {journal} {\bibinfo  {journal} {Nature Photonics}\ }\textbf {\bibinfo {volume} {15}},\ \bibinfo {pages} {499--503} (\bibinfo {year} {2021})}\BibitemShut {NoStop}%
\bibitem [{\citenamefont {Maznev}\ \emph {et~al.}(2018)\citenamefont {Maznev}, \citenamefont {Bencivenga}, \citenamefont {Cannizzo}, \citenamefont {Capotondi}, \citenamefont {Cucini}, \citenamefont {Duncan}, \citenamefont {Feurer}, \citenamefont {Frazer}, \citenamefont {Foglia}, \citenamefont {Frey} \emph {et~al.}}]{maznev2018generation}%
  \BibitemOpen
  \bibfield  {author} {\bibinfo {author} {\bibfnamefont {A.}~\bibnamefont {Maznev}}, \bibinfo {author} {\bibfnamefont {F.}~\bibnamefont {Bencivenga}}, \bibinfo {author} {\bibfnamefont {A.}~\bibnamefont {Cannizzo}}, \bibinfo {author} {\bibfnamefont {F.}~\bibnamefont {Capotondi}}, \bibinfo {author} {\bibfnamefont {R.}~\bibnamefont {Cucini}}, \bibinfo {author} {\bibfnamefont {R.}~\bibnamefont {Duncan}}, \bibinfo {author} {\bibfnamefont {T.}~\bibnamefont {Feurer}}, \bibinfo {author} {\bibfnamefont {T.}~\bibnamefont {Frazer}}, \bibinfo {author} {\bibfnamefont {L.}~\bibnamefont {Foglia}}, \bibinfo {author} {\bibfnamefont {H.-M.}\ \bibnamefont {Frey}}, \emph {et~al.},\ }\bibfield  {title} {\enquote {\bibinfo {title} {Generation of coherent phonons by coherent extreme ultraviolet radiation in a transient grating experiment},}\ }\href@noop {} {\bibfield  {journal} {\bibinfo  {journal} {Applied Physics Letters}\ }\textbf {\bibinfo {volume} {113}},\ \bibinfo {pages} {221905} (\bibinfo {year} {2018})}\BibitemShut
  {NoStop}%
\bibitem [{\citenamefont {Maznev}\ \emph {et~al.}(2021)\citenamefont {Maznev}, \citenamefont {Mincigrucci}, \citenamefont {Bencivenga}, \citenamefont {Unikandanunni}, \citenamefont {Capotondi}, \citenamefont {Chen}, \citenamefont {Ding}, \citenamefont {Duncan}, \citenamefont {Foglia}, \citenamefont {Izzo} \emph {et~al.}}]{maznev2021generation}%
  \BibitemOpen
  \bibfield  {author} {\bibinfo {author} {\bibfnamefont {A.}~\bibnamefont {Maznev}}, \bibinfo {author} {\bibfnamefont {R.}~\bibnamefont {Mincigrucci}}, \bibinfo {author} {\bibfnamefont {F.}~\bibnamefont {Bencivenga}}, \bibinfo {author} {\bibfnamefont {V.}~\bibnamefont {Unikandanunni}}, \bibinfo {author} {\bibfnamefont {F.}~\bibnamefont {Capotondi}}, \bibinfo {author} {\bibfnamefont {G.}~\bibnamefont {Chen}}, \bibinfo {author} {\bibfnamefont {Z.}~\bibnamefont {Ding}}, \bibinfo {author} {\bibfnamefont {R.}~\bibnamefont {Duncan}}, \bibinfo {author} {\bibfnamefont {L.}~\bibnamefont {Foglia}}, \bibinfo {author} {\bibfnamefont {M.}~\bibnamefont {Izzo}}, \emph {et~al.},\ }\bibfield  {title} {\enquote {\bibinfo {title} {Generation and detection of 50 ghz surface acoustic waves by extreme ultraviolet pulses},}\ }\href@noop {} {\bibfield  {journal} {\bibinfo  {journal} {Applied Physics Letters}\ }\textbf {\bibinfo {volume} {119}},\ \bibinfo {pages} {044102} (\bibinfo {year} {2021})}\BibitemShut {NoStop}%
\bibitem [{\citenamefont {Milloch}\ \emph {et~al.}(2021)\citenamefont {Milloch}, \citenamefont {Mincigrucci}, \citenamefont {Capotondi}, \citenamefont {De~Angelis}, \citenamefont {Foglia}, \citenamefont {Kurdi}, \citenamefont {Naumenko}, \citenamefont {Pedersoli}, \citenamefont {Pelli-Cresi}, \citenamefont {Simoncig} \emph {et~al.}}]{milloch2021nanoscale}%
  \BibitemOpen
  \bibfield  {author} {\bibinfo {author} {\bibfnamefont {A.}~\bibnamefont {Milloch}}, \bibinfo {author} {\bibfnamefont {R.}~\bibnamefont {Mincigrucci}}, \bibinfo {author} {\bibfnamefont {F.}~\bibnamefont {Capotondi}}, \bibinfo {author} {\bibfnamefont {D.}~\bibnamefont {De~Angelis}}, \bibinfo {author} {\bibfnamefont {L.}~\bibnamefont {Foglia}}, \bibinfo {author} {\bibfnamefont {G.}~\bibnamefont {Kurdi}}, \bibinfo {author} {\bibfnamefont {D.}~\bibnamefont {Naumenko}}, \bibinfo {author} {\bibfnamefont {E.}~\bibnamefont {Pedersoli}}, \bibinfo {author} {\bibfnamefont {J.~S.}\ \bibnamefont {Pelli-Cresi}}, \bibinfo {author} {\bibfnamefont {A.}~\bibnamefont {Simoncig}}, \emph {et~al.},\ }\bibfield  {title} {\enquote {\bibinfo {title} {Nanoscale thermoelasticity in silicon nitride membranes: Implications for thermal management},}\ }\href@noop {} {\bibfield  {journal} {\bibinfo  {journal} {ACS Applied Nano Materials}\ }\textbf {\bibinfo {volume} {4}},\ \bibinfo {pages} {10519--10527} (\bibinfo {year}
  {2021})}\BibitemShut {NoStop}%
\bibitem [{\citenamefont {Svetina}\ \emph {et~al.}(2019)\citenamefont {Svetina}, \citenamefont {Mankowsky}, \citenamefont {Knopp}, \citenamefont {Koch}, \citenamefont {Seniutinas}, \citenamefont {R{\"o}sner}, \citenamefont {Kubec}, \citenamefont {Lebugle}, \citenamefont {Mochi}, \citenamefont {Beck} \emph {et~al.}}]{svetina2019towards}%
  \BibitemOpen
  \bibfield  {author} {\bibinfo {author} {\bibfnamefont {C.}~\bibnamefont {Svetina}}, \bibinfo {author} {\bibfnamefont {R.}~\bibnamefont {Mankowsky}}, \bibinfo {author} {\bibfnamefont {G.}~\bibnamefont {Knopp}}, \bibinfo {author} {\bibfnamefont {F.}~\bibnamefont {Koch}}, \bibinfo {author} {\bibfnamefont {G.}~\bibnamefont {Seniutinas}}, \bibinfo {author} {\bibfnamefont {B.}~\bibnamefont {R{\"o}sner}}, \bibinfo {author} {\bibfnamefont {A.}~\bibnamefont {Kubec}}, \bibinfo {author} {\bibfnamefont {M.}~\bibnamefont {Lebugle}}, \bibinfo {author} {\bibfnamefont {I.}~\bibnamefont {Mochi}}, \bibinfo {author} {\bibfnamefont {M.}~\bibnamefont {Beck}}, \emph {et~al.},\ }\bibfield  {title} {\enquote {\bibinfo {title} {Towards x-ray transient grating spectroscopy},}\ }\href@noop {} {\bibfield  {journal} {\bibinfo  {journal} {Optics Letters}\ }\textbf {\bibinfo {volume} {44}},\ \bibinfo {pages} {574--577} (\bibinfo {year} {2019})}\BibitemShut {NoStop}%
\bibitem [{\citenamefont {Weder}\ \emph {et~al.}(2020)\citenamefont {Weder}, \citenamefont {von Korff~Schmising}, \citenamefont {G{\"u}nther}, \citenamefont {Schneider}, \citenamefont {Engel}, \citenamefont {Hessing}, \citenamefont {Str{\"u}ber}, \citenamefont {Weigand}, \citenamefont {Vodungbo}, \citenamefont {Jal} \emph {et~al.}}]{weder2020transient}%
  \BibitemOpen
  \bibfield  {author} {\bibinfo {author} {\bibfnamefont {D.}~\bibnamefont {Weder}}, \bibinfo {author} {\bibfnamefont {C.}~\bibnamefont {von Korff~Schmising}}, \bibinfo {author} {\bibfnamefont {C.}~\bibnamefont {G{\"u}nther}}, \bibinfo {author} {\bibfnamefont {M.}~\bibnamefont {Schneider}}, \bibinfo {author} {\bibfnamefont {D.}~\bibnamefont {Engel}}, \bibinfo {author} {\bibfnamefont {P.}~\bibnamefont {Hessing}}, \bibinfo {author} {\bibfnamefont {C.}~\bibnamefont {Str{\"u}ber}}, \bibinfo {author} {\bibfnamefont {M.}~\bibnamefont {Weigand}}, \bibinfo {author} {\bibfnamefont {B.}~\bibnamefont {Vodungbo}}, \bibinfo {author} {\bibfnamefont {E.}~\bibnamefont {Jal}}, \emph {et~al.},\ }\bibfield  {title} {\enquote {\bibinfo {title} {Transient magnetic gratings on the nanometer scale},}\ }\href@noop {} {\bibfield  {journal} {\bibinfo  {journal} {Structural Dynamics}\ }\textbf {\bibinfo {volume} {7}},\ \bibinfo {pages} {054501} (\bibinfo {year} {2020})}\BibitemShut {NoStop}%
\bibitem [{\citenamefont {Ksenzov}\ \emph {et~al.}(2021)\citenamefont {Ksenzov}, \citenamefont {Maznev}, \citenamefont {Unikandanunni}, \citenamefont {Bencivenga}, \citenamefont {Capotondi}, \citenamefont {Caretta}, \citenamefont {Foglia}, \citenamefont {Malvestuto}, \citenamefont {Masciovecchio}, \citenamefont {Mincigrucci} \emph {et~al.}}]{ksenzov2021nanoscale}%
  \BibitemOpen
  \bibfield  {author} {\bibinfo {author} {\bibfnamefont {D.}~\bibnamefont {Ksenzov}}, \bibinfo {author} {\bibfnamefont {A.~A.}\ \bibnamefont {Maznev}}, \bibinfo {author} {\bibfnamefont {V.}~\bibnamefont {Unikandanunni}}, \bibinfo {author} {\bibfnamefont {F.}~\bibnamefont {Bencivenga}}, \bibinfo {author} {\bibfnamefont {F.}~\bibnamefont {Capotondi}}, \bibinfo {author} {\bibfnamefont {A.}~\bibnamefont {Caretta}}, \bibinfo {author} {\bibfnamefont {L.}~\bibnamefont {Foglia}}, \bibinfo {author} {\bibfnamefont {M.}~\bibnamefont {Malvestuto}}, \bibinfo {author} {\bibfnamefont {C.}~\bibnamefont {Masciovecchio}}, \bibinfo {author} {\bibfnamefont {R.}~\bibnamefont {Mincigrucci}}, \emph {et~al.},\ }\bibfield  {title} {\enquote {\bibinfo {title} {Nanoscale transient magnetization gratings created and probed by femtosecond extreme ultraviolet pulses},}\ }\href@noop {} {\bibfield  {journal} {\bibinfo  {journal} {Nano Letters}\ }\textbf {\bibinfo {volume} {21}},\ \bibinfo {pages} {2905--2911} (\bibinfo {year}
  {2021})}\BibitemShut {NoStop}%
\bibitem [{\citenamefont {Yao}\ \emph {et~al.}(2022)\citenamefont {Yao}, \citenamefont {Steinbach}, \citenamefont {Borchert}, \citenamefont {Schick}, \citenamefont {Engel}, \citenamefont {Bencivenga}, \citenamefont {Mincigrucci}, \citenamefont {Foglia}, \citenamefont {Pedersoli}, \citenamefont {De~Angelis} \emph {et~al.}}]{yao2022all}%
  \BibitemOpen
  \bibfield  {author} {\bibinfo {author} {\bibfnamefont {K.}~\bibnamefont {Yao}}, \bibinfo {author} {\bibfnamefont {F.}~\bibnamefont {Steinbach}}, \bibinfo {author} {\bibfnamefont {M.}~\bibnamefont {Borchert}}, \bibinfo {author} {\bibfnamefont {D.}~\bibnamefont {Schick}}, \bibinfo {author} {\bibfnamefont {D.}~\bibnamefont {Engel}}, \bibinfo {author} {\bibfnamefont {F.}~\bibnamefont {Bencivenga}}, \bibinfo {author} {\bibfnamefont {R.}~\bibnamefont {Mincigrucci}}, \bibinfo {author} {\bibfnamefont {L.}~\bibnamefont {Foglia}}, \bibinfo {author} {\bibfnamefont {E.}~\bibnamefont {Pedersoli}}, \bibinfo {author} {\bibfnamefont {D.}~\bibnamefont {De~Angelis}}, \emph {et~al.},\ }\bibfield  {title} {\enquote {\bibinfo {title} {All-optical switching on the nanometer scale excited and probed with femtosecond extreme ultraviolet pulses},}\ }\href@noop {} {\bibfield  {journal} {\bibinfo  {journal} {Nano Letters}\ }\textbf {\bibinfo {volume} {22}},\ \bibinfo {pages} {4452–4458} (\bibinfo {year} {2022})}\BibitemShut
  {NoStop}%
\bibitem [{\citenamefont {Eichler}, \citenamefont {G{\"u}nter},\ and\ \citenamefont {Pohl}(1986)}]{eichler2013laser}%
  \BibitemOpen
  \bibfield  {author} {\bibinfo {author} {\bibfnamefont {H.~J.}\ \bibnamefont {Eichler}}, \bibinfo {author} {\bibfnamefont {P.}~\bibnamefont {G{\"u}nter}},\ and\ \bibinfo {author} {\bibfnamefont {D.~W.}\ \bibnamefont {Pohl}},\ }\href@noop {} {\emph {\bibinfo {title} {Laser-induced dynamic gratings}}}\ (\bibinfo  {publisher} {Springer-Verlag},\ \bibinfo {address} {Berlin},\ \bibinfo {year} {1986})\BibitemShut {NoStop}%
\bibitem [{\citenamefont {Janu{\v{s}}onis}\ \emph {et~al.}(2016)\citenamefont {Janu{\v{s}}onis}, \citenamefont {Jansma}, \citenamefont {Chang}, \citenamefont {Liu}, \citenamefont {Gatilova}, \citenamefont {Lomonosov}, \citenamefont {Shalagatskyi}, \citenamefont {Pezeril}, \citenamefont {Temnov},\ and\ \citenamefont {Tobey}}]{januvsonis2016transient}%
  \BibitemOpen
  \bibfield  {author} {\bibinfo {author} {\bibfnamefont {J.}~\bibnamefont {Janu{\v{s}}onis}}, \bibinfo {author} {\bibfnamefont {T.}~\bibnamefont {Jansma}}, \bibinfo {author} {\bibfnamefont {C.}~\bibnamefont {Chang}}, \bibinfo {author} {\bibfnamefont {Q.}~\bibnamefont {Liu}}, \bibinfo {author} {\bibfnamefont {A.}~\bibnamefont {Gatilova}}, \bibinfo {author} {\bibfnamefont {A.}~\bibnamefont {Lomonosov}}, \bibinfo {author} {\bibfnamefont {V.}~\bibnamefont {Shalagatskyi}}, \bibinfo {author} {\bibfnamefont {T.}~\bibnamefont {Pezeril}}, \bibinfo {author} {\bibfnamefont {V.}~\bibnamefont {Temnov}},\ and\ \bibinfo {author} {\bibfnamefont {R.}~\bibnamefont {Tobey}},\ }\bibfield  {title} {\enquote {\bibinfo {title} {Transient grating spectroscopy in magnetic thin films: Simultaneous detection of elastic and magnetic dynamics},}\ }\href@noop {} {\bibfield  {journal} {\bibinfo  {journal} {Scientific Reports}\ }\textbf {\bibinfo {volume} {6}},\ \bibinfo {pages} {1--10} (\bibinfo {year} {2016})}\BibitemShut {NoStop}%
\bibitem [{\citenamefont {Coey}(2020)}]{coey2020perspective}%
  \BibitemOpen
  \bibfield  {author} {\bibinfo {author} {\bibfnamefont {J.}~\bibnamefont {Coey}},\ }\bibfield  {title} {\enquote {\bibinfo {title} {Perspective and prospects for rare earth permanent magnets},}\ }\href@noop {} {\bibfield  {journal} {\bibinfo  {journal} {Engineering}\ }\textbf {\bibinfo {volume} {6}},\ \bibinfo {pages} {119--131} (\bibinfo {year} {2020})}\BibitemShut {NoStop}%
\bibitem [{\citenamefont {Gonz{\'a}lez}, \citenamefont {Andr{\'e}s},\ and\ \citenamefont {L{\'o}pez~Ant{\'o}n}(2021)}]{gonzalez2021applied}%
  \BibitemOpen
  \bibfield  {author} {\bibinfo {author} {\bibfnamefont {J.~A.}\ \bibnamefont {Gonz{\'a}lez}}, \bibinfo {author} {\bibfnamefont {J.~P.}\ \bibnamefont {Andr{\'e}s}},\ and\ \bibinfo {author} {\bibfnamefont {R.}~\bibnamefont {L{\'o}pez~Ant{\'o}n}},\ }\bibfield  {title} {\enquote {\bibinfo {title} {Applied trends in magnetic rare earth/transition metal alloys and multilayers},}\ }\href@noop {} {\bibfield  {journal} {\bibinfo  {journal} {Sensors}\ }\textbf {\bibinfo {volume} {21}},\ \bibinfo {pages} {5615} (\bibinfo {year} {2021})}\BibitemShut {NoStop}%
\bibitem [{\citenamefont {Hansen}\ \emph {et~al.}(1991)\citenamefont {Hansen}, \citenamefont {Klahn}, \citenamefont {Clausen}, \citenamefont {Much},\ and\ \citenamefont {Witter}}]{hansen1991magnetic}%
  \BibitemOpen
  \bibfield  {author} {\bibinfo {author} {\bibfnamefont {P.}~\bibnamefont {Hansen}}, \bibinfo {author} {\bibfnamefont {S.}~\bibnamefont {Klahn}}, \bibinfo {author} {\bibfnamefont {C.}~\bibnamefont {Clausen}}, \bibinfo {author} {\bibfnamefont {G.}~\bibnamefont {Much}},\ and\ \bibinfo {author} {\bibfnamefont {K.}~\bibnamefont {Witter}},\ }\bibfield  {title} {\enquote {\bibinfo {title} {Magnetic and magneto-optical properties of rare-earth transition-metal alloys containing {Dy, Ho, Fe, Co}},}\ }\href@noop {} {\bibfield  {journal} {\bibinfo  {journal} {Journal of Applied Physics}\ }\textbf {\bibinfo {volume} {69}},\ \bibinfo {pages} {3194--3207} (\bibinfo {year} {1991})}\BibitemShut {NoStop}%
\bibitem [{\citenamefont {Andreev}\ and\ \citenamefont {Zadvorkin}(1991)}]{andreev1991thermal}%
  \BibitemOpen
  \bibfield  {author} {\bibinfo {author} {\bibfnamefont {A.}~\bibnamefont {Andreev}}\ and\ \bibinfo {author} {\bibfnamefont {S.}~\bibnamefont {Zadvorkin}},\ }\bibfield  {title} {\enquote {\bibinfo {title} {Thermal expansion and spontaneous magnetostriction of {RCo$_5$} intermetallic compounds},}\ }\href@noop {} {\bibfield  {journal} {\bibinfo  {journal} {Physica B: Condensed Matter}\ }\textbf {\bibinfo {volume} {172}},\ \bibinfo {pages} {517--525} (\bibinfo {year} {1991})}\BibitemShut {NoStop}%
\bibitem [{\citenamefont {Radwanski}(1987)}]{radwanski1987origin}%
  \BibitemOpen
  \bibfield  {author} {\bibinfo {author} {\bibfnamefont {R.}~\bibnamefont {Radwanski}},\ }\bibfield  {title} {\enquote {\bibinfo {title} {The origin of the basal-plane magnetocrystalline anisotropy in 4f {Co-rich} intermetallics},}\ }\href@noop {} {\bibfield  {journal} {\bibinfo  {journal} {Journal of Physics F: Metal Physics}\ }\textbf {\bibinfo {volume} {17}},\ \bibinfo {pages} {267} (\bibinfo {year} {1987})}\BibitemShut {NoStop}%
\bibitem [{\citenamefont {Tie-Song}\ \emph {et~al.}(1991)\citenamefont {Tie-Song}, \citenamefont {Han-Min}, \citenamefont {Guang-Hua}, \citenamefont {Xiu-Feng},\ and\ \citenamefont {Hong}}]{tie1991magnetic}%
  \BibitemOpen
  \bibfield  {author} {\bibinfo {author} {\bibfnamefont {Z.}~\bibnamefont {Tie-Song}}, \bibinfo {author} {\bibfnamefont {J.}~\bibnamefont {Han-Min}}, \bibinfo {author} {\bibfnamefont {G.}~\bibnamefont {Guang-Hua}}, \bibinfo {author} {\bibfnamefont {H.}~\bibnamefont {Xiu-Feng}},\ and\ \bibinfo {author} {\bibfnamefont {C.}~\bibnamefont {Hong}},\ }\bibfield  {title} {\enquote {\bibinfo {title} {Magnetic properties of {R} ions in {RCo$_5$} compounds ({R=Pr, Nd, Sm, Gd, Tb, Dy, Ho, and Er})},}\ }\href@noop {} {\bibfield  {journal} {\bibinfo  {journal} {Physical Review B}\ }\textbf {\bibinfo {volume} {43}},\ \bibinfo {pages} {8593} (\bibinfo {year} {1991})}\BibitemShut {NoStop}%
\bibitem [{\citenamefont {Donges}\ \emph {et~al.}(2017)\citenamefont {Donges}, \citenamefont {Khmelevskyi}, \citenamefont {Deak}, \citenamefont {Abrudan}, \citenamefont {Schmitz}, \citenamefont {Radu}, \citenamefont {Radu}, \citenamefont {Szunyogh},\ and\ \citenamefont {Nowak}}]{donges2017magnetization}%
  \BibitemOpen
  \bibfield  {author} {\bibinfo {author} {\bibfnamefont {A.}~\bibnamefont {Donges}}, \bibinfo {author} {\bibfnamefont {S.}~\bibnamefont {Khmelevskyi}}, \bibinfo {author} {\bibfnamefont {A.}~\bibnamefont {Deak}}, \bibinfo {author} {\bibfnamefont {R.-M.}\ \bibnamefont {Abrudan}}, \bibinfo {author} {\bibfnamefont {D.}~\bibnamefont {Schmitz}}, \bibinfo {author} {\bibfnamefont {I.}~\bibnamefont {Radu}}, \bibinfo {author} {\bibfnamefont {F.}~\bibnamefont {Radu}}, \bibinfo {author} {\bibfnamefont {L.}~\bibnamefont {Szunyogh}},\ and\ \bibinfo {author} {\bibfnamefont {U.}~\bibnamefont {Nowak}},\ }\bibfield  {title} {\enquote {\bibinfo {title} {Magnetization compensation and spin reorientation transition in ferrimagnetic {DyCo$_5$}: Multiscale modeling and element-specific measurements},}\ }\href@noop {} {\bibfield  {journal} {\bibinfo  {journal} {Physical Review B}\ }\textbf {\bibinfo {volume} {96}},\ \bibinfo {pages} {024412} (\bibinfo {year} {2017})}\BibitemShut {NoStop}%
\bibitem [{\citenamefont {Grechishkin}\ \emph {et~al.}(2017)\citenamefont {Grechishkin}, \citenamefont {Ivanova}, \citenamefont {Grachev}, \citenamefont {Zigert},\ and\ \citenamefont {Eguzhokova}}]{grechishkin2017domain}%
  \BibitemOpen
  \bibfield  {author} {\bibinfo {author} {\bibfnamefont {R.}~\bibnamefont {Grechishkin}}, \bibinfo {author} {\bibfnamefont {A.}~\bibnamefont {Ivanova}}, \bibinfo {author} {\bibfnamefont {A.}~\bibnamefont {Grachev}}, \bibinfo {author} {\bibfnamefont {A.}~\bibnamefont {Zigert}},\ and\ \bibinfo {author} {\bibfnamefont {R.}~\bibnamefont {Eguzhokova}},\ }\bibfield  {title} {\enquote {\bibinfo {title} {Domain structure and spin reorientation in {TbCo$_5$ and DyCo$_5$} intermetallics},}\ }\href@noop {} {\bibfield  {journal} {\bibinfo  {journal} {IEEE Transactions on Magnetics}\ }\textbf {\bibinfo {volume} {53}},\ \bibinfo {pages} {1--4} (\bibinfo {year} {2017})}\BibitemShut {NoStop}%
\bibitem [{\citenamefont {Kirilyuk}, \citenamefont {Kimel},\ and\ \citenamefont {Rasing}(2010)}]{kirilyuk2010ultrafast}%
  \BibitemOpen
  \bibfield  {author} {\bibinfo {author} {\bibfnamefont {A.}~\bibnamefont {Kirilyuk}}, \bibinfo {author} {\bibfnamefont {A.~V.}\ \bibnamefont {Kimel}},\ and\ \bibinfo {author} {\bibfnamefont {T.}~\bibnamefont {Rasing}},\ }\bibfield  {title} {\enquote {\bibinfo {title} {Ultrafast optical manipulation of magnetic order},}\ }\href@noop {} {\bibfield  {journal} {\bibinfo  {journal} {Reviews of Modern Physics}\ }\textbf {\bibinfo {volume} {82}},\ \bibinfo {pages} {2731} (\bibinfo {year} {2010})}\BibitemShut {NoStop}%
\bibitem [{\citenamefont {Radu}\ \emph {et~al.}(2011)\citenamefont {Radu}, \citenamefont {Vahaplar}, \citenamefont {Stamm}, \citenamefont {Kachel}, \citenamefont {Pontius}, \citenamefont {D{\"u}rr}, \citenamefont {Ostler}, \citenamefont {Barker}, \citenamefont {Evans}, \citenamefont {Chantrell} \emph {et~al.}}]{radu2011transient}%
  \BibitemOpen
  \bibfield  {author} {\bibinfo {author} {\bibfnamefont {I.}~\bibnamefont {Radu}}, \bibinfo {author} {\bibfnamefont {K.}~\bibnamefont {Vahaplar}}, \bibinfo {author} {\bibfnamefont {C.}~\bibnamefont {Stamm}}, \bibinfo {author} {\bibfnamefont {T.}~\bibnamefont {Kachel}}, \bibinfo {author} {\bibfnamefont {N.}~\bibnamefont {Pontius}}, \bibinfo {author} {\bibfnamefont {H.}~\bibnamefont {D{\"u}rr}}, \bibinfo {author} {\bibfnamefont {T.}~\bibnamefont {Ostler}}, \bibinfo {author} {\bibfnamefont {J.}~\bibnamefont {Barker}}, \bibinfo {author} {\bibfnamefont {R.}~\bibnamefont {Evans}}, \bibinfo {author} {\bibfnamefont {R.}~\bibnamefont {Chantrell}}, \emph {et~al.},\ }\bibfield  {title} {\enquote {\bibinfo {title} {Transient ferromagnetic-like state mediating ultrafast reversal of antiferromagnetically coupled spins},}\ }\href@noop {} {\bibfield  {journal} {\bibinfo  {journal} {Nature}\ }\textbf {\bibinfo {volume} {472}},\ \bibinfo {pages} {205--208} (\bibinfo {year} {2011})}\BibitemShut {NoStop}%
\bibitem [{\citenamefont {Radu}\ \emph {et~al.}(2012)\citenamefont {Radu}, \citenamefont {Abrudan}, \citenamefont {Radu}, \citenamefont {Schmitz},\ and\ \citenamefont {Zabel}}]{radu2012perpendicular}%
  \BibitemOpen
  \bibfield  {author} {\bibinfo {author} {\bibfnamefont {F.}~\bibnamefont {Radu}}, \bibinfo {author} {\bibfnamefont {R.}~\bibnamefont {Abrudan}}, \bibinfo {author} {\bibfnamefont {I.}~\bibnamefont {Radu}}, \bibinfo {author} {\bibfnamefont {D.}~\bibnamefont {Schmitz}},\ and\ \bibinfo {author} {\bibfnamefont {H.}~\bibnamefont {Zabel}},\ }\bibfield  {title} {\enquote {\bibinfo {title} {Perpendicular exchange bias in ferrimagnetic spin valves},}\ }\href@noop {} {\bibfield  {journal} {\bibinfo  {journal} {Nature Communications}\ }\textbf {\bibinfo {volume} {3}},\ \bibinfo {pages} {1--7} (\bibinfo {year} {2012})}\BibitemShut {NoStop}%
\bibitem [{\citenamefont {{\"U}nal}\ \emph {et~al.}(2016)\citenamefont {{\"U}nal}, \citenamefont {Valencia}, \citenamefont {Radu}, \citenamefont {Marchenko}, \citenamefont {Merazzo}, \citenamefont {V{\'a}zquez},\ and\ \citenamefont {S{\'a}nchez-Barriga}}]{unal2016ferrimagnetic}%
  \BibitemOpen
  \bibfield  {author} {\bibinfo {author} {\bibfnamefont {A.}~\bibnamefont {{\"U}nal}}, \bibinfo {author} {\bibfnamefont {S.}~\bibnamefont {Valencia}}, \bibinfo {author} {\bibfnamefont {F.}~\bibnamefont {Radu}}, \bibinfo {author} {\bibfnamefont {D.}~\bibnamefont {Marchenko}}, \bibinfo {author} {\bibfnamefont {K.}~\bibnamefont {Merazzo}}, \bibinfo {author} {\bibfnamefont {M.}~\bibnamefont {V{\'a}zquez}},\ and\ \bibinfo {author} {\bibfnamefont {J.}~\bibnamefont {S{\'a}nchez-Barriga}},\ }\bibfield  {title} {\enquote {\bibinfo {title} {Ferrimagnetic {DyCo$_5$} nanostructures for bits in heat-assisted magnetic recording},}\ }\href@noop {} {\bibfield  {journal} {\bibinfo  {journal} {Physical Review Applied}\ }\textbf {\bibinfo {volume} {5}},\ \bibinfo {pages} {064007} (\bibinfo {year} {2016})}\BibitemShut {NoStop}%
\bibitem [{\citenamefont {Radu}\ and\ \citenamefont {S{\'a}nchez-Barriga}(2018)}]{radu2018ferrimagnetic}%
  \BibitemOpen
  \bibfield  {author} {\bibinfo {author} {\bibfnamefont {F.}~\bibnamefont {Radu}}\ and\ \bibinfo {author} {\bibfnamefont {J.}~\bibnamefont {S{\'a}nchez-Barriga}},\ }\bibfield  {title} {\enquote {\bibinfo {title} {Ferrimagnetic heterostructures for applications in magnetic recording},}\ }in\ \href@noop {} {\emph {\bibinfo {booktitle} {Novel Magnetic Nanostructures}}}\ (\bibinfo  {publisher} {Elsevier},\ \bibinfo {year} {2018})\ pp.\ \bibinfo {pages} {267--331}\BibitemShut {NoStop}%
\bibitem [{\citenamefont {Chen}\ \emph {et~al.}(2020)\citenamefont {Chen}, \citenamefont {Lott}, \citenamefont {Philippi-Kobs}, \citenamefont {Weigand}, \citenamefont {Luo},\ and\ \citenamefont {Radu}}]{chen2020observation}%
  \BibitemOpen
  \bibfield  {author} {\bibinfo {author} {\bibfnamefont {K.}~\bibnamefont {Chen}}, \bibinfo {author} {\bibfnamefont {D.}~\bibnamefont {Lott}}, \bibinfo {author} {\bibfnamefont {A.}~\bibnamefont {Philippi-Kobs}}, \bibinfo {author} {\bibfnamefont {M.}~\bibnamefont {Weigand}}, \bibinfo {author} {\bibfnamefont {C.}~\bibnamefont {Luo}},\ and\ \bibinfo {author} {\bibfnamefont {F.}~\bibnamefont {Radu}},\ }\bibfield  {title} {\enquote {\bibinfo {title} {Observation of compact ferrimagnetic skyrmions in {DyCo$_3$} film},}\ }\href@noop {} {\bibfield  {journal} {\bibinfo  {journal} {Nanoscale}\ }\textbf {\bibinfo {volume} {12}},\ \bibinfo {pages} {18137--18143} (\bibinfo {year} {2020})}\BibitemShut {NoStop}%
\bibitem [{\citenamefont {Seifert}\ \emph {et~al.}(2021)\citenamefont {Seifert}, \citenamefont {Martens}, \citenamefont {Radu}, \citenamefont {Ribow}, \citenamefont {Berritta}, \citenamefont {N{\'a}dvorn{\'\i}k}, \citenamefont {Starke}, \citenamefont {Jungwirth}, \citenamefont {Wolf}, \citenamefont {Radu} \emph {et~al.}}]{seifert2021frequency}%
  \BibitemOpen
  \bibfield  {author} {\bibinfo {author} {\bibfnamefont {T.~S.}\ \bibnamefont {Seifert}}, \bibinfo {author} {\bibfnamefont {U.}~\bibnamefont {Martens}}, \bibinfo {author} {\bibfnamefont {F.}~\bibnamefont {Radu}}, \bibinfo {author} {\bibfnamefont {M.}~\bibnamefont {Ribow}}, \bibinfo {author} {\bibfnamefont {M.}~\bibnamefont {Berritta}}, \bibinfo {author} {\bibfnamefont {L.}~\bibnamefont {N{\'a}dvorn{\'\i}k}}, \bibinfo {author} {\bibfnamefont {R.}~\bibnamefont {Starke}}, \bibinfo {author} {\bibfnamefont {T.}~\bibnamefont {Jungwirth}}, \bibinfo {author} {\bibfnamefont {M.}~\bibnamefont {Wolf}}, \bibinfo {author} {\bibfnamefont {I.}~\bibnamefont {Radu}}, \emph {et~al.},\ }\bibfield  {title} {\enquote {\bibinfo {title} {Frequency-independent terahertz anomalous hall effect in {DyCo$_5$, Co$_{32}$Fe$_{68}$, and Gd$_27$Fe$_{73}$} thin films from {DC} to 40 {THz}},}\ }\href@noop {} {\bibfield  {journal} {\bibinfo  {journal} {Advanced Materials}\ }\textbf {\bibinfo {volume} {33}},\ \bibinfo {pages} {2007398} (\bibinfo
  {year} {2021})}\BibitemShut {NoStop}%
\bibitem [{\citenamefont {del Moral}, \citenamefont {Algarabel},\ and\ \citenamefont {Ibarra}(1987)}]{del1987magnetoelastic}%
  \BibitemOpen
  \bibfield  {author} {\bibinfo {author} {\bibfnamefont {A.}~\bibnamefont {del Moral}}, \bibinfo {author} {\bibfnamefont {P.}~\bibnamefont {Algarabel}},\ and\ \bibinfo {author} {\bibfnamefont {M.}~\bibnamefont {Ibarra}},\ }\bibfield  {title} {\enquote {\bibinfo {title} {Magnetoelastic coupling and spin reorientation in {RECo$_5$} uniaxial magnets ({RE= Pr, Dy, Ho and Y}). {II}},}\ }\href@noop {} {\bibfield  {journal} {\bibinfo  {journal} {Journal of Magnetism and Magnetic Materials}\ }\textbf {\bibinfo {volume} {69}},\ \bibinfo {pages} {285--298} (\bibinfo {year} {1987})}\BibitemShut {NoStop}%
\bibitem [{\citenamefont {Kamar{\'a}d}, \citenamefont {Arnold},\ and\ \citenamefont {Ibarra}(1995)}]{kamarad1995magnetic}%
  \BibitemOpen
  \bibfield  {author} {\bibinfo {author} {\bibfnamefont {J.}~\bibnamefont {Kamar{\'a}d}}, \bibinfo {author} {\bibfnamefont {Z.}~\bibnamefont {Arnold}},\ and\ \bibinfo {author} {\bibfnamefont {M.}~\bibnamefont {Ibarra}},\ }\bibfield  {title} {\enquote {\bibinfo {title} {Magnetic phase transitions and magnetovolume anomalies in {DyCo$_2$} and {GdMn$_2$} compounds under pressure},}\ }\href@noop {} {\bibfield  {journal} {\bibinfo  {journal} {Journal of Magnetism and Magnetic Materials}\ }\textbf {\bibinfo {volume} {140}},\ \bibinfo {pages} {837--838} (\bibinfo {year} {1995})}\BibitemShut {NoStop}%
\bibitem [{\citenamefont {Puebla}\ \emph {et~al.}(2022)\citenamefont {Puebla}, \citenamefont {Hwang}, \citenamefont {Maekawa},\ and\ \citenamefont {Otani}}]{puebla2022perspectives}%
  \BibitemOpen
  \bibfield  {author} {\bibinfo {author} {\bibfnamefont {J.}~\bibnamefont {Puebla}}, \bibinfo {author} {\bibfnamefont {Y.}~\bibnamefont {Hwang}}, \bibinfo {author} {\bibfnamefont {S.}~\bibnamefont {Maekawa}},\ and\ \bibinfo {author} {\bibfnamefont {Y.}~\bibnamefont {Otani}},\ }\bibfield  {title} {\enquote {\bibinfo {title} {Perspectives on spintronics with surface acoustic waves},}\ }\href@noop {} {\bibfield  {journal} {\bibinfo  {journal} {Applied Physics Letters}\ }\textbf {\bibinfo {volume} {120}},\ \bibinfo {pages} {220502} (\bibinfo {year} {2022})}\BibitemShut {NoStop}%
\bibitem [{\citenamefont {Luo}\ \emph {et~al.}(2019)\citenamefont {Luo}, \citenamefont {Ryll}, \citenamefont {Back},\ and\ \citenamefont {Radu}}]{luo2019x}%
  \BibitemOpen
  \bibfield  {author} {\bibinfo {author} {\bibfnamefont {C.}~\bibnamefont {Luo}}, \bibinfo {author} {\bibfnamefont {H.}~\bibnamefont {Ryll}}, \bibinfo {author} {\bibfnamefont {C.~H.}\ \bibnamefont {Back}},\ and\ \bibinfo {author} {\bibfnamefont {F.}~\bibnamefont {Radu}},\ }\bibfield  {title} {\enquote {\bibinfo {title} {X-ray magnetic linear dichroism as a probe for non-collinear magnetic state in ferrimagnetic single layer exchange bias systems},}\ }\href@noop {} {\bibfield  {journal} {\bibinfo  {journal} {Scientific Reports}\ }\textbf {\bibinfo {volume} {9}},\ \bibinfo {pages} {1--9} (\bibinfo {year} {2019})}\BibitemShut {NoStop}%
\bibitem [{\citenamefont {Glavic}\ and\ \citenamefont {Bj{\"o}rck}(2022)}]{glavic2022genx}%
  \BibitemOpen
  \bibfield  {author} {\bibinfo {author} {\bibfnamefont {A.}~\bibnamefont {Glavic}}\ and\ \bibinfo {author} {\bibfnamefont {M.}~\bibnamefont {Bj{\"o}rck}},\ }\bibfield  {title} {\enquote {\bibinfo {title} {Genx 3: the latest generation of an established tool},}\ }\href@noop {} {\bibfield  {journal} {\bibinfo  {journal} {Journal of Applied Crystallography}\ }\textbf {\bibinfo {volume} {55}} (\bibinfo {year} {2022})}\BibitemShut {NoStop}%
\bibitem [{\citenamefont {Daillant}\ and\ \citenamefont {Gibaud}(2008)}]{daillant2008x}%
  \BibitemOpen
  \bibfield  {author} {\bibinfo {author} {\bibfnamefont {J.}~\bibnamefont {Daillant}}\ and\ \bibinfo {author} {\bibfnamefont {A.}~\bibnamefont {Gibaud}},\ }\href@noop {} {\emph {\bibinfo {title} {X-ray and neutron reflectivity: principles and applications}}},\ Vol.\ \bibinfo {volume} {770}\ (\bibinfo  {publisher} {Springer},\ \bibinfo {year} {2008})\BibitemShut {NoStop}%
\bibitem [{\citenamefont {St{\"o}hr}\ and\ \citenamefont {Siegmann}(2006)}]{stohr2006magnetism}%
  \BibitemOpen
  \bibfield  {author} {\bibinfo {author} {\bibfnamefont {J.}~\bibnamefont {St{\"o}hr}}\ and\ \bibinfo {author} {\bibfnamefont {H.~C.}\ \bibnamefont {Siegmann}},\ }\bibfield  {title} {\enquote {\bibinfo {title} {Magnetism},}\ }\href@noop {} {\bibfield  {journal} {\bibinfo  {journal} {Solid-State Sciences. Springer, Berlin, Heidelberg}\ }\textbf {\bibinfo {volume} {5}},\ \bibinfo {pages} {236} (\bibinfo {year} {2006})}\BibitemShut {NoStop}%
\bibitem [{\citenamefont {Noll}, \citenamefont {Radu}\ \emph {et~al.}(2016)\citenamefont {Noll}, \citenamefont {Radu} \emph {et~al.}}]{noll2016mechanics}%
  \BibitemOpen
  \bibfield  {author} {\bibinfo {author} {\bibfnamefont {T.}~\bibnamefont {Noll}}, \bibinfo {author} {\bibfnamefont {F.}~\bibnamefont {Radu}}, \emph {et~al.},\ }\bibfield  {title} {\enquote {\bibinfo {title} {The mechanics of the {VEKMAG} experiment},}\ }\href@noop {} {\bibfield  {journal} {\bibinfo  {journal} {Proc. of MEDSI2016, Barcelona, Spain}\ ,\ \bibinfo {pages} {370--373}} (\bibinfo {year} {2016})}\BibitemShut {NoStop}%
\bibitem [{\citenamefont {Mincigrucci}\ \emph {et~al.}(2018)\citenamefont {Mincigrucci}, \citenamefont {Foglia}, \citenamefont {Naumenko}, \citenamefont {Pedersoli}, \citenamefont {Simoncig}, \citenamefont {Cucini}, \citenamefont {Gessini}, \citenamefont {Kiskinova}, \citenamefont {Kurdi}, \citenamefont {Mahne} \emph {et~al.}}]{mincigrucci2018advances}%
  \BibitemOpen
  \bibfield  {author} {\bibinfo {author} {\bibfnamefont {R.}~\bibnamefont {Mincigrucci}}, \bibinfo {author} {\bibfnamefont {L.}~\bibnamefont {Foglia}}, \bibinfo {author} {\bibfnamefont {D.}~\bibnamefont {Naumenko}}, \bibinfo {author} {\bibfnamefont {E.}~\bibnamefont {Pedersoli}}, \bibinfo {author} {\bibfnamefont {A.}~\bibnamefont {Simoncig}}, \bibinfo {author} {\bibfnamefont {R.}~\bibnamefont {Cucini}}, \bibinfo {author} {\bibfnamefont {A.}~\bibnamefont {Gessini}}, \bibinfo {author} {\bibfnamefont {M.}~\bibnamefont {Kiskinova}}, \bibinfo {author} {\bibfnamefont {G.}~\bibnamefont {Kurdi}}, \bibinfo {author} {\bibfnamefont {N.}~\bibnamefont {Mahne}}, \emph {et~al.},\ }\bibfield  {title} {\enquote {\bibinfo {title} {Advances in instrumentation for {FEL}-based four-wave-mixing experiments},}\ }\href@noop {} {\bibfield  {journal} {\bibinfo  {journal} {Nuclear Instruments and Methods in Physics Research Section A: Accelerators, Spectrometers, Detectors and Associated Equipment}\ }\textbf {\bibinfo {volume} {907}},\
  \bibinfo {pages} {132--148} (\bibinfo {year} {2018})}\BibitemShut {NoStop}%
\bibitem [{\citenamefont {Paolasini}(2014)}]{paolasini2014resonant}%
  \BibitemOpen
  \bibfield  {author} {\bibinfo {author} {\bibfnamefont {L.}~\bibnamefont {Paolasini}},\ }\bibfield  {title} {\enquote {\bibinfo {title} {Resonant and magnetic x-ray diffraction by polarized synchrotron radiation},}\ }\href@noop {} {\bibfield  {journal} {\bibinfo  {journal} {{\'E}cole th{\'e}matique de la Soci{\'e}t{\'e} Fran{\c{c}}aise de la Neutronique}\ }\textbf {\bibinfo {volume} {13}},\ \bibinfo {pages} {03002} (\bibinfo {year} {2014})}\BibitemShut {NoStop}%
\bibitem [{\citenamefont {Ukleev}\ \emph {et~al.}(2023)\citenamefont {Ukleev}, \citenamefont {Burian}, \citenamefont {Gliga}, \citenamefont {Vaz}, \citenamefont {R{\"o}sner}, \citenamefont {Fainozzi}, \citenamefont {Seniutinas}, \citenamefont {Kubec}, \citenamefont {Mankowsky}, \citenamefont {Lemke} \emph {et~al.}}]{ukleev2023effect}%
  \BibitemOpen
  \bibfield  {author} {\bibinfo {author} {\bibfnamefont {V.}~\bibnamefont {Ukleev}}, \bibinfo {author} {\bibfnamefont {M.}~\bibnamefont {Burian}}, \bibinfo {author} {\bibfnamefont {S.}~\bibnamefont {Gliga}}, \bibinfo {author} {\bibfnamefont {C.}~\bibnamefont {Vaz}}, \bibinfo {author} {\bibfnamefont {B.}~\bibnamefont {R{\"o}sner}}, \bibinfo {author} {\bibfnamefont {D.}~\bibnamefont {Fainozzi}}, \bibinfo {author} {\bibfnamefont {G.}~\bibnamefont {Seniutinas}}, \bibinfo {author} {\bibfnamefont {A.}~\bibnamefont {Kubec}}, \bibinfo {author} {\bibfnamefont {R.}~\bibnamefont {Mankowsky}}, \bibinfo {author} {\bibfnamefont {H.~T.}\ \bibnamefont {Lemke}}, \emph {et~al.},\ }\bibfield  {title} {\enquote {\bibinfo {title} {Effect of intense x-ray free-electron laser transient gratings on the magnetic domain structure of {Tm: YIG}},}\ }\href@noop {} {\bibfield  {journal} {\bibinfo  {journal} {Journal of Applied Physics}\ }\textbf {\bibinfo {volume} {133}},\ \bibinfo {pages} {123902} (\bibinfo {year} {2023})}\BibitemShut
  {NoStop}%
\bibitem [{\citenamefont {Naumenko}\ \emph {et~al.}(2019)\citenamefont {Naumenko}, \citenamefont {Mincigrucci}, \citenamefont {Altissimo}, \citenamefont {Foglia}, \citenamefont {Gessini}, \citenamefont {Kurdi}, \citenamefont {Nikolov}, \citenamefont {Pedersoli}, \citenamefont {Principi}, \citenamefont {Simoncig} \emph {et~al.}}]{naumenko2019thermoelasticity}%
  \BibitemOpen
  \bibfield  {author} {\bibinfo {author} {\bibfnamefont {D.}~\bibnamefont {Naumenko}}, \bibinfo {author} {\bibfnamefont {R.}~\bibnamefont {Mincigrucci}}, \bibinfo {author} {\bibfnamefont {M.}~\bibnamefont {Altissimo}}, \bibinfo {author} {\bibfnamefont {L.}~\bibnamefont {Foglia}}, \bibinfo {author} {\bibfnamefont {A.}~\bibnamefont {Gessini}}, \bibinfo {author} {\bibfnamefont {G.}~\bibnamefont {Kurdi}}, \bibinfo {author} {\bibfnamefont {I.}~\bibnamefont {Nikolov}}, \bibinfo {author} {\bibfnamefont {E.}~\bibnamefont {Pedersoli}}, \bibinfo {author} {\bibfnamefont {E.}~\bibnamefont {Principi}}, \bibinfo {author} {\bibfnamefont {A.}~\bibnamefont {Simoncig}}, \emph {et~al.},\ }\bibfield  {title} {\enquote {\bibinfo {title} {Thermoelasticity of nanoscale silicon carbide membranes excited by extreme ultraviolet transient gratings: implications for mechanical and thermal management},}\ }\href@noop {} {\bibfield  {journal} {\bibinfo  {journal} {ACS Applied Nano Materials}\ }\textbf {\bibinfo {volume} {2}},\ \bibinfo
  {pages} {5132--5139} (\bibinfo {year} {2019})}\BibitemShut {NoStop}%
\bibitem [{\citenamefont {Rogers}\ \emph {et~al.}(2000)\citenamefont {Rogers}, \citenamefont {Maznev}, \citenamefont {Banet},\ and\ \citenamefont {Nelson}}]{rogers2000optical}%
  \BibitemOpen
  \bibfield  {author} {\bibinfo {author} {\bibfnamefont {J.~A.}\ \bibnamefont {Rogers}}, \bibinfo {author} {\bibfnamefont {A.~A.}\ \bibnamefont {Maznev}}, \bibinfo {author} {\bibfnamefont {M.~J.}\ \bibnamefont {Banet}},\ and\ \bibinfo {author} {\bibfnamefont {K.~A.}\ \bibnamefont {Nelson}},\ }\bibfield  {title} {\enquote {\bibinfo {title} {Optical generation and characterization of acoustic waves in thin films: Fundamental and applications},}\ }\href@noop {} {\bibfield  {journal} {\bibinfo  {journal} {Annual Review of Materials Research}\ }\textbf {\bibinfo {volume} {30}},\ \bibinfo {pages} {117} (\bibinfo {year} {2000})}\BibitemShut {NoStop}%
\bibitem [{\citenamefont {Zhou}\ \emph {et~al.}(2014)\citenamefont {Zhou}, \citenamefont {Talbi}, \citenamefont {Tiercelin},\ and\ \citenamefont {Bou~Matar}}]{zhou2014multilayer}%
  \BibitemOpen
  \bibfield  {author} {\bibinfo {author} {\bibfnamefont {H.}~\bibnamefont {Zhou}}, \bibinfo {author} {\bibfnamefont {A.}~\bibnamefont {Talbi}}, \bibinfo {author} {\bibfnamefont {N.}~\bibnamefont {Tiercelin}},\ and\ \bibinfo {author} {\bibfnamefont {O.}~\bibnamefont {Bou~Matar}},\ }\bibfield  {title} {\enquote {\bibinfo {title} {Multilayer magnetostrictive structure based surface acoustic wave devices},}\ }\href@noop {} {\bibfield  {journal} {\bibinfo  {journal} {Applied Physics Letters}\ }\textbf {\bibinfo {volume} {104}} (\bibinfo {year} {2014})}\BibitemShut {NoStop}%
\bibitem [{\citenamefont {K{\"a}ding}\ \emph {et~al.}(1995)\citenamefont {K{\"a}ding}, \citenamefont {Skurk}, \citenamefont {Maznev},\ and\ \citenamefont {Matthias}}]{kading1995transient}%
  \BibitemOpen
  \bibfield  {author} {\bibinfo {author} {\bibfnamefont {O.}~\bibnamefont {K{\"a}ding}}, \bibinfo {author} {\bibfnamefont {H.}~\bibnamefont {Skurk}}, \bibinfo {author} {\bibfnamefont {A.}~\bibnamefont {Maznev}},\ and\ \bibinfo {author} {\bibfnamefont {E.}~\bibnamefont {Matthias}},\ }\bibfield  {title} {\enquote {\bibinfo {title} {Transient thermal gratings at surfaces for thermal characterization of bulk materials and thin films},}\ }\href@noop {} {\bibfield  {journal} {\bibinfo  {journal} {Applied Physics A}\ }\textbf {\bibinfo {volume} {61}},\ \bibinfo {pages} {253--261} (\bibinfo {year} {1995})}\BibitemShut {NoStop}%
\bibitem [{\citenamefont {Abrudan}\ \emph {et~al.}(2021)\citenamefont {Abrudan}, \citenamefont {Hennecke}, \citenamefont {Radu}, \citenamefont {Kachel}, \citenamefont {Holldack}, \citenamefont {Mitzner}, \citenamefont {Donges}, \citenamefont {Khmelevskyi}, \citenamefont {De{\'a}k}, \citenamefont {Szunyogh} \emph {et~al.}}]{abrudan2021element}%
  \BibitemOpen
  \bibfield  {author} {\bibinfo {author} {\bibfnamefont {R.}~\bibnamefont {Abrudan}}, \bibinfo {author} {\bibfnamefont {M.}~\bibnamefont {Hennecke}}, \bibinfo {author} {\bibfnamefont {F.}~\bibnamefont {Radu}}, \bibinfo {author} {\bibfnamefont {T.}~\bibnamefont {Kachel}}, \bibinfo {author} {\bibfnamefont {K.}~\bibnamefont {Holldack}}, \bibinfo {author} {\bibfnamefont {R.}~\bibnamefont {Mitzner}}, \bibinfo {author} {\bibfnamefont {A.}~\bibnamefont {Donges}}, \bibinfo {author} {\bibfnamefont {S.}~\bibnamefont {Khmelevskyi}}, \bibinfo {author} {\bibfnamefont {A.}~\bibnamefont {De{\'a}k}}, \bibinfo {author} {\bibfnamefont {L.}~\bibnamefont {Szunyogh}}, \emph {et~al.},\ }\bibfield  {title} {\enquote {\bibinfo {title} {Element-specific magnetization damping in ferrimagnetic {DyCo$_5$} alloys revealed by ultrafast x-ray measurements},}\ }\href@noop {} {\bibfield  {journal} {\bibinfo  {journal} {Physica Status Solidi (RRL)--Rapid Research Letters}\ }\textbf {\bibinfo {volume} {15}},\ \bibinfo {pages} {2100047}
  (\bibinfo {year} {2021})}\BibitemShut {NoStop}%
\bibitem [{\citenamefont {Bencivenga}\ \emph {et~al.}(2023)\citenamefont {Bencivenga}, \citenamefont {Capotondi}, \citenamefont {Foglia}, \citenamefont {Mincigrucci},\ and\ \citenamefont {Masciovecchio}}]{bencivenga2023extreme}%
  \BibitemOpen
  \bibfield  {author} {\bibinfo {author} {\bibfnamefont {F.}~\bibnamefont {Bencivenga}}, \bibinfo {author} {\bibfnamefont {F.}~\bibnamefont {Capotondi}}, \bibinfo {author} {\bibfnamefont {L.}~\bibnamefont {Foglia}}, \bibinfo {author} {\bibfnamefont {R.}~\bibnamefont {Mincigrucci}},\ and\ \bibinfo {author} {\bibfnamefont {C.}~\bibnamefont {Masciovecchio}},\ }\bibfield  {title} {\enquote {\bibinfo {title} {Extreme ultraviolet transient gratings},}\ }\href@noop {} {\bibfield  {journal} {\bibinfo  {journal} {Advances in Physics: X}\ }\textbf {\bibinfo {volume} {8}},\ \bibinfo {pages} {2220363} (\bibinfo {year} {2023})}\BibitemShut {NoStop}%
\bibitem [{\citenamefont {Thompson}, \citenamefont {Vaughan}\ \emph {et~al.}(2001)\citenamefont {Thompson}, \citenamefont {Vaughan} \emph {et~al.}}]{thompson2001x}%
  \BibitemOpen
  \bibfield  {author} {\bibinfo {author} {\bibfnamefont {A.~C.}\ \bibnamefont {Thompson}}, \bibinfo {author} {\bibfnamefont {D.}~\bibnamefont {Vaughan}}, \emph {et~al.},\ }\href@noop {} {\emph {\bibinfo {title} {X-ray data booklet}}},\ Vol.~\bibinfo {volume} {8}\ (\bibinfo  {publisher} {Lawrence Berkeley National Laboratory, University of California Berkeley, CA},\ \bibinfo {year} {2001})\BibitemShut {NoStop}%
\bibitem [{\citenamefont {Foglia}\ \emph {et~al.}(2023{\natexlab{a}})\citenamefont {Foglia}, \citenamefont {Wehinger}, \citenamefont {Perosa}, \citenamefont {Mincigrucci}, \citenamefont {Allaria}, \citenamefont {Armillotta}, \citenamefont {Brynes}, \citenamefont {Cucini}, \citenamefont {De~Angelis}, \citenamefont {De~Ninno} \emph {et~al.}}]{foglia2023nanoscale}%
  \BibitemOpen
  \bibfield  {author} {\bibinfo {author} {\bibfnamefont {L.}~\bibnamefont {Foglia}}, \bibinfo {author} {\bibfnamefont {B.}~\bibnamefont {Wehinger}}, \bibinfo {author} {\bibfnamefont {G.}~\bibnamefont {Perosa}}, \bibinfo {author} {\bibfnamefont {R.}~\bibnamefont {Mincigrucci}}, \bibinfo {author} {\bibfnamefont {E.}~\bibnamefont {Allaria}}, \bibinfo {author} {\bibfnamefont {F.}~\bibnamefont {Armillotta}}, \bibinfo {author} {\bibfnamefont {A.}~\bibnamefont {Brynes}}, \bibinfo {author} {\bibfnamefont {R.}~\bibnamefont {Cucini}}, \bibinfo {author} {\bibfnamefont {D.}~\bibnamefont {De~Angelis}}, \bibinfo {author} {\bibfnamefont {G.}~\bibnamefont {De~Ninno}}, \emph {et~al.},\ }\bibfield  {title} {\enquote {\bibinfo {title} {Nanoscale transient polarization gratings},}\ }\href@noop {} {\bibfield  {journal} {\bibinfo  {journal} {arXiv preprint arXiv:2310.15734}\ } (\bibinfo {year} {2023}{\natexlab{a}})}\BibitemShut {NoStop}%
\bibitem [{\citenamefont {Foglia}\ \emph {et~al.}(2023{\natexlab{b}})\citenamefont {Foglia}, \citenamefont {Mincigrucci}, \citenamefont {Maznev}, \citenamefont {Baldi}, \citenamefont {Capotondi}, \citenamefont {Caporaletti}, \citenamefont {Comin}, \citenamefont {De~Angelis}, \citenamefont {Duncan}, \citenamefont {Fainozzi} \emph {et~al.}}]{foglia2023extreme}%
  \BibitemOpen
  \bibfield  {author} {\bibinfo {author} {\bibfnamefont {L.}~\bibnamefont {Foglia}}, \bibinfo {author} {\bibfnamefont {R.}~\bibnamefont {Mincigrucci}}, \bibinfo {author} {\bibfnamefont {A.}~\bibnamefont {Maznev}}, \bibinfo {author} {\bibfnamefont {G.}~\bibnamefont {Baldi}}, \bibinfo {author} {\bibfnamefont {F.}~\bibnamefont {Capotondi}}, \bibinfo {author} {\bibfnamefont {F.}~\bibnamefont {Caporaletti}}, \bibinfo {author} {\bibfnamefont {R.}~\bibnamefont {Comin}}, \bibinfo {author} {\bibfnamefont {D.}~\bibnamefont {De~Angelis}}, \bibinfo {author} {\bibfnamefont {R.}~\bibnamefont {Duncan}}, \bibinfo {author} {\bibfnamefont {D.}~\bibnamefont {Fainozzi}}, \emph {et~al.},\ }\bibfield  {title} {\enquote {\bibinfo {title} {Extreme ultraviolet transient gratings: A tool for nanoscale photoacoustics},}\ }\href@noop {} {\bibfield  {journal} {\bibinfo  {journal} {Photoacoustics}\ }\textbf {\bibinfo {volume} {29}},\ \bibinfo {pages} {100453} (\bibinfo {year} {2023}{\natexlab{b}})}\BibitemShut {NoStop}%
\bibitem [{\citenamefont {Yokouchi}\ \emph {et~al.}(2020)\citenamefont {Yokouchi}, \citenamefont {Sugimoto}, \citenamefont {Rana}, \citenamefont {Seki}, \citenamefont {Ogawa}, \citenamefont {Kasai},\ and\ \citenamefont {Otani}}]{yokouchi2020creation}%
  \BibitemOpen
  \bibfield  {author} {\bibinfo {author} {\bibfnamefont {T.}~\bibnamefont {Yokouchi}}, \bibinfo {author} {\bibfnamefont {S.}~\bibnamefont {Sugimoto}}, \bibinfo {author} {\bibfnamefont {B.}~\bibnamefont {Rana}}, \bibinfo {author} {\bibfnamefont {S.}~\bibnamefont {Seki}}, \bibinfo {author} {\bibfnamefont {N.}~\bibnamefont {Ogawa}}, \bibinfo {author} {\bibfnamefont {S.}~\bibnamefont {Kasai}},\ and\ \bibinfo {author} {\bibfnamefont {Y.}~\bibnamefont {Otani}},\ }\bibfield  {title} {\enquote {\bibinfo {title} {Creation of magnetic skyrmions by surface acoustic waves},}\ }\href@noop {} {\bibfield  {journal} {\bibinfo  {journal} {Nature Nanotechnology}\ }\textbf {\bibinfo {volume} {15}},\ \bibinfo {pages} {361--366} (\bibinfo {year} {2020})}\BibitemShut {NoStop}%
\bibitem [{\citenamefont {Chen}\ \emph {et~al.}(2021)\citenamefont {Chen}, \citenamefont {Lin}, \citenamefont {Niu}, \citenamefont {Sun}, \citenamefont {Yang}, \citenamefont {Kang},\ and\ \citenamefont {Lei}}]{chen2021surface}%
  \BibitemOpen
  \bibfield  {author} {\bibinfo {author} {\bibfnamefont {C.}~\bibnamefont {Chen}}, \bibinfo {author} {\bibfnamefont {T.}~\bibnamefont {Lin}}, \bibinfo {author} {\bibfnamefont {J.}~\bibnamefont {Niu}}, \bibinfo {author} {\bibfnamefont {Y.}~\bibnamefont {Sun}}, \bibinfo {author} {\bibfnamefont {L.}~\bibnamefont {Yang}}, \bibinfo {author} {\bibfnamefont {W.}~\bibnamefont {Kang}},\ and\ \bibinfo {author} {\bibfnamefont {N.}~\bibnamefont {Lei}},\ }\bibfield  {title} {\enquote {\bibinfo {title} {Surface acoustic wave controlled skyrmion-based synapse devices},}\ }\href@noop {} {\bibfield  {journal} {\bibinfo  {journal} {Nanotechnology}\ }\textbf {\bibinfo {volume} {33}},\ \bibinfo {pages} {115205} (\bibinfo {year} {2021})}\BibitemShut {NoStop}%
\bibitem [{\citenamefont {Kobayashi}\ \emph {et~al.}(2017)\citenamefont {Kobayashi}, \citenamefont {Yoshikawa}, \citenamefont {Matsuo}, \citenamefont {Iguchi}, \citenamefont {Maekawa}, \citenamefont {Saitoh},\ and\ \citenamefont {Nozaki}}]{kobayashi2017spin}%
  \BibitemOpen
  \bibfield  {author} {\bibinfo {author} {\bibfnamefont {D.}~\bibnamefont {Kobayashi}}, \bibinfo {author} {\bibfnamefont {T.}~\bibnamefont {Yoshikawa}}, \bibinfo {author} {\bibfnamefont {M.}~\bibnamefont {Matsuo}}, \bibinfo {author} {\bibfnamefont {R.}~\bibnamefont {Iguchi}}, \bibinfo {author} {\bibfnamefont {S.}~\bibnamefont {Maekawa}}, \bibinfo {author} {\bibfnamefont {E.}~\bibnamefont {Saitoh}},\ and\ \bibinfo {author} {\bibfnamefont {Y.}~\bibnamefont {Nozaki}},\ }\bibfield  {title} {\enquote {\bibinfo {title} {Spin current generation using a surface acoustic wave generated via spin-rotation coupling},}\ }\href@noop {} {\bibfield  {journal} {\bibinfo  {journal} {Physical Review Letters}\ }\textbf {\bibinfo {volume} {119}},\ \bibinfo {pages} {077202} (\bibinfo {year} {2017})}\BibitemShut {NoStop}%
\end{thebibliography}%
\end{document}